\documentclass[twocolumn,english,nobibnotes,nofootinbib,superscriptaddress ]{revtex4-1}
\usepackage[T1]{fontenc}
\usepackage[latin9]{inputenc}
\usepackage{geometry}
\usepackage{amsbsy}
\usepackage{graphicx}
\usepackage{babel}
\usepackage{color}
\usepackage{amsmath}
\begin{document}

\author{E.N. Osika}
\affiliation{AGH University of Science and Technology, Faculty of Physics and
Applied Computer Science,\\
 al. Mickiewicza 30, 30-059 Krak\'ow, Poland}

\author{A. Chac\'on}
\affiliation{ICFO  -  Institut  de  Ciencies  Fotoniques,
The  Barcelona  Institute  of  Science  and  Technology,
08860  Castelldefels  (Barcelona),  Spain}

\author{M. Lewenstein}
\affiliation{ICFO  -  Institut  de  Ciencies  Fotoniques,
The  Barcelona  Institute  of  Science  and  Technology,
08860  Castelldefels  (Barcelona),  Spain}
\affiliation{ICREA, Pg. Llu\'is Companys 23, 08010 Barcelona, Spain}

\author{B. Szafran}
\affiliation{AGH University of Science and Technology, Faculty of Physics and
Applied Computer Science,\\
 al. Mickiewicza 30, 30-059 Krak\'ow, Poland}

\title{Spin-valley dynamics of electrically driven ambipolar carbon-nanotube quantum dots }
\begin{abstract}
An ambipolar $n$-$p$ double quantum dot defined by potential variation along a semiconducting carbon-nanotube is considered. We focus on the (1e,1h) charge configuration with a single excess electron in the conduction band state confined in the $n$-type dot and a single missing electron in the valence band state of the   $p$-dot for which  lifting of the Pauli blockade of the current was observed in the electric-dipole spin resonance [E. A. Laird et al. Nat. Nanotech. 8 , 565 (2013)].
The dynamics of the system driven by periodic electric field is studied with the Floquet theory and the time-dependent configuration interaction method with 
the single-electron spin-valley-orbitals determined for atomistic tight-binding Hamiltonian.   We find that the transitions lifting the Pauli blockade 
are strongly influenced by coupling to a vacuum state with an empty $n$ dot and a fully filled $p$ dot. 
The coupling shifts the transition energies and strongly modifies the effective $g$ factors for axial magnetic field. The coupling is modulated by the bias between the dots but it appears effective for surprisingly large energy splitting
between the (1e,1h) ground state and the vacuum (0e,0h) state.
 Multiphoton transitions and high harmonic generation effects are also discussed.
\end{abstract}
\maketitle
\section{Introduction}


Manipulation of the spin degree of freedom for electrons confined in quantum dots (QDs) has been under extensive
studies in the context of construction of spintronic single-electron devices \cite{fabian} for over a decade. 
For QDs  a successful  implementation of the electron spin-resonance was performed  with a microwave generator integrated into the device \cite{koppens}.
The magnetic field produced by ac currents \cite{koppens} was soon replaced  by the effective magnetic field due to the spin-orbit coupling for a driven electron motion within the QD \cite{nadji,extreme,stroer},
in the electric-dipole spin resonance \cite{edsr} (EDSR). EDSR can also be induced by fluctuations of the Overhauser field \cite{laird} or inhomogeneous field \cite{weg} translated into an effective ac magnetic field averaged by the wave function of a periodically driven electron \cite{osi1}.   The detection of the spin flip \cite{koppens,extreme,stroer,nadji,laird} exploits the Pauli blockade \cite{pauliblockade} of the current that flows across a double QD.

The lifting of the Pauli blockade induced by ac electric field has been observed in double QDs defined within a semiconducting carbon nanotube (CNT) \cite{pei,lairdpei}, where
the electron dynamics involves both the spin and the valley \cite{pabu} degree of freedom. The EDSR was observed for double QDs in an ambipolar work point: with one quantum dot storing an extra electron of the conduction band and the other a hole (a single-unoccupied state) in the valence band -- denoted as (1e,1h) in the following. 
The energy spectrum of the (1e,1h) system was determined  with the atomistic tight binding approach  in Ref. \cite{eo_bent}. The driven electron dynamics was discussed in Ref. \cite{li} in a continuum approach strictly in the subspace corresponding to the (1e,1h) charge configuration. The dynamics of the (1e,1e) system in double carbon nanotube unipolar dots has also been considered \cite{leakage}.

The quantum dot confined single-electron states are nearly fourfold degenerate with a pair of spin-valley doublets split by spin-orbit coupling energy of the order of 1 meV \cite{kum},
so that the dynamics of the system in the experimental work point \cite{pei,lairdpei} involves a single electron in the $n$ dot and three electrons in the $p$ dot. 
In this work we solve the problem of the spin-valley transitions between the Pauli blocked and nonblocked states using a time dependent configuration interaction approach and Floquet theory \cite{Shirley, Chu} for the ambipolar dots. We find that the states of the (1e,1h) charge configuration are strongly coupled by the ac potential with the ''vacuum state'' (0e,0h) -- with an empty $n$-dot and fully filled $p$-dot,  even when its energy is relatively high above the (1e,1h) ground state. The coupling -- beyond the subspace considered in Ref. \cite{li} -- produces a strong shift of the transition lines off the energy spectra, which for axial magnetic field strongly modifies the effective $g$ factors for the driven spin-valley transitions. 

The present approach -- that provides exact result for the coherent few-electron dynamics including the higher order effects  --  besides energy shifts describes  also  the multiphoton transitions \cite{pei,extreme,subh,danon}. The dynamics of QDs in external ac voltages enters the regime of strongly driven systems already at relatively weak voltages. Motivated by this fact we look for the high harmonic generation (HHG) effects that 
\begin{figure}[htbp]
\includegraphics[width=0.95\linewidth]{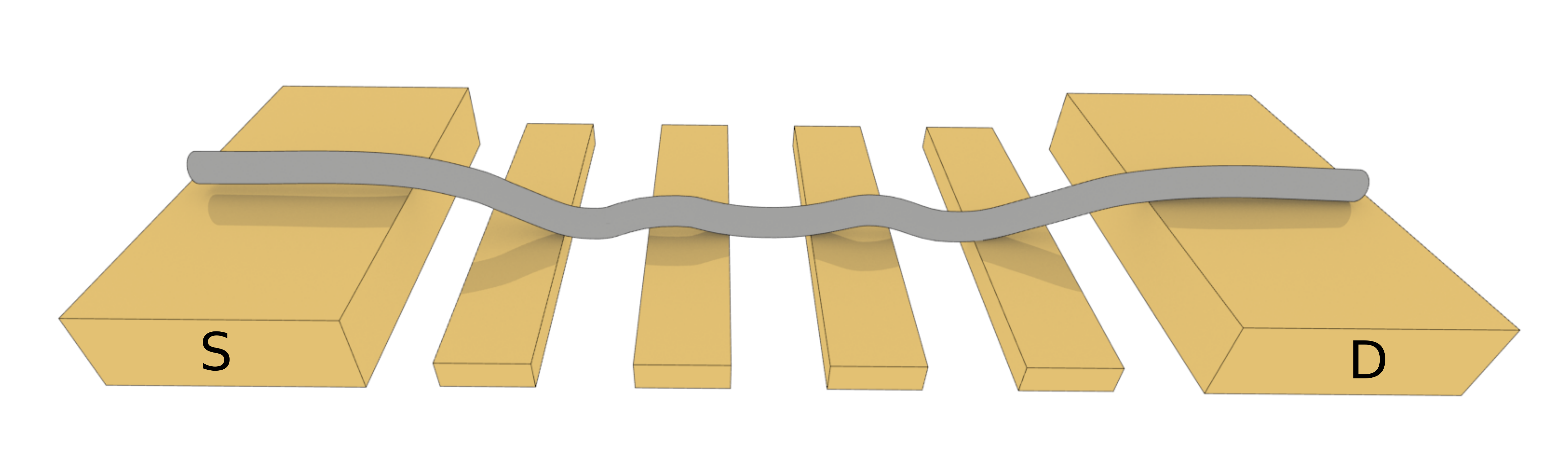}
\caption{ Schematics of the considered system -- carbon nanotube deflected above electrostatic gates forming the QD confinement potential.
} \label{nanotube}
\end{figure}
are encountered in systems driven by strong laser fields  (for HHG from atomic or molecular sources see \cite{lewen,atto},   for recent studies of HHG in solids see \cite{aso,solid-hhg, icfo-hhg,ciappina-rev}). Higher harmonics of the driven dipole moment are found but only in resonant conditions.

\begin{figure*}[htbp]
\includegraphics[width=0.7\linewidth]{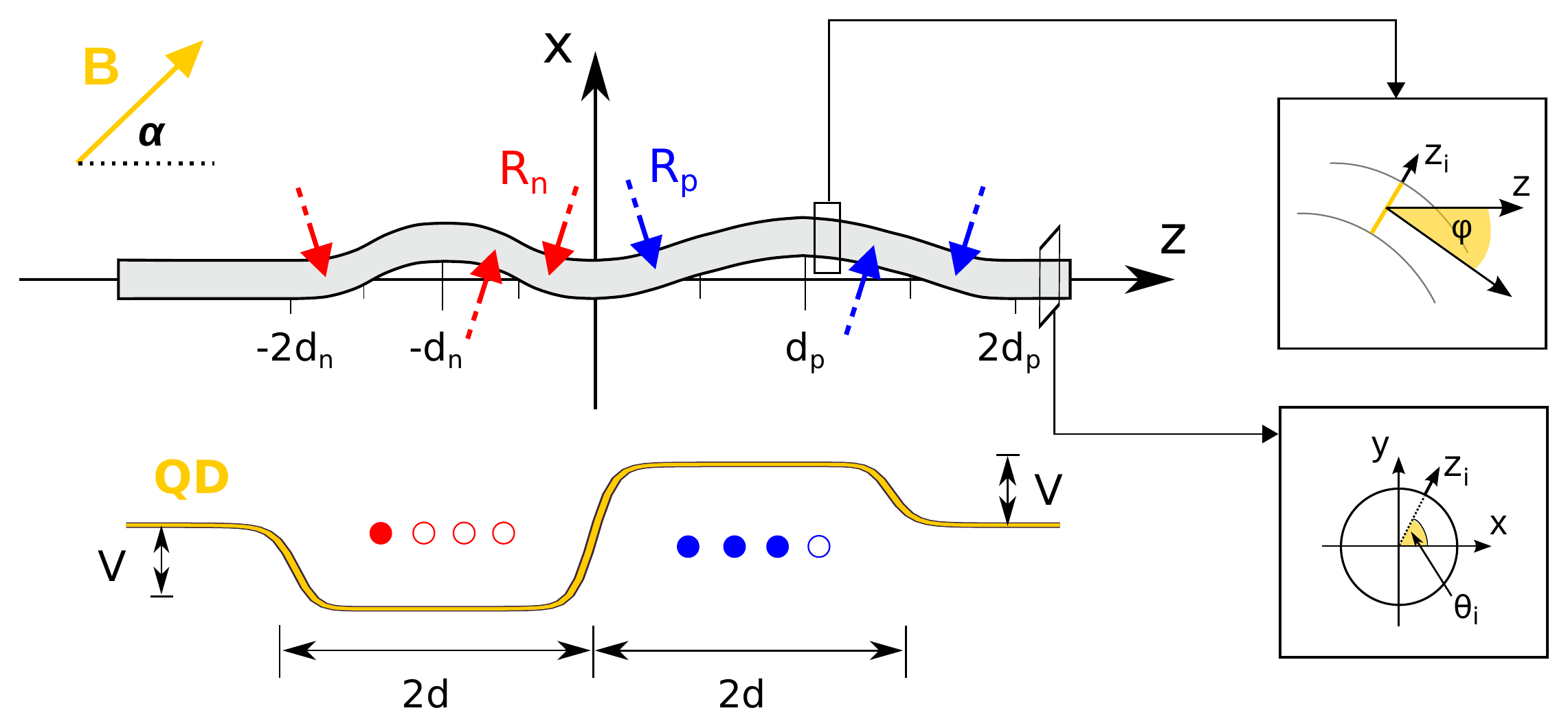}
\caption{ Schematics of the modeled CNT with all the dimensions and coordinate systems used in the calculations explained.  
} \label{schemat}
\end{figure*}

\section{Theory}

\subsection{Model} 
In this section we first discuss a model, describing a $n-p$ quantum dot induced within a carbon nanotube. In the next section (B) we present the method that we use to describe the dynamics of the system: time-dependent configuration interaction approach. Finally, in section C we discuss the Floquet approach, specifically suited for the treatment of the time-dependent problems with periodic time modulations.

We consider a carbon nanotube of length $L=53.11$ nm, diameter $2r=1.33$ nm, with the chiral vector $C_{h}=(17,0)$ for which the CNT is semiconducting and 
can confine electrons in quantum dots defined electrostatically.
The nanotube is considered bent \cite{flensb} above the electrostatic gates as in Fig. \ref{nanotube} and in the experiment [see Fig. 2(a) of Ref. \onlinecite{lairdpei}].  
We have found \cite{eo_bent} that the bends of the CNT appearing at the area where the confinement potential is defined on the gates
results in energy spectra which qualitatively agree with the experimental spectrum of transitions lifting the Pauli spin-valley blockade of the current \cite{lairdpei}.

The global $z$ axis of the chosen coordinate system coincides with the axis of the straight part of the CNT axis [Fig. \ref{schemat}]. 
The nanotube is deflected within the global $xz$-plane.
The radii of the bents in Fig. \ref{schemat} are $R_n=8$ nm and $R_p=11$ nm. The positions of the centre of the arcs along the $z$ direction are $d_n=8$ nm and $d_p=11$ nm, respectively. 
Locally the bend is parametrized by the inclination angle $\phi_{i}$ of CNT axis (for $i$-th ion) to the global $z$ axis [see the inset to Fig. \ref{schemat}].

We model a double $n-p$ quantum dot induced within the nanotube by external voltages. The shape of the electrostatic confinement potential is described by 
\begin{equation}
W_{QD}(z)=-\frac{V_{n}}{1+(\frac{z+z_{s}}{d})^{14}}+\frac{V_{p}}{1+(\frac{z-z_{s}}{d})^{14}},
\end{equation}
where $V_n$ and $V_p$ are potentials on the $n$ and $p$ dots, respectively, $z_s$ is a shift of the QD centre from $z=0$ and $d$ is a half width of the QD. In the calculations we use $V_n=V_p=0.21$ eV, $z_s=8$ nm and $2d=16$ nm. The experimental data \cite{lairdpei} contain signatures of the intervalley mixing. We account for the intervalley scattering
introducing a potential peak of 1 eV at a single  atom at $z=-12.28$ nm which is responsible for the mixing of the $K$ and $K'$ orbital states.
In part of the calculations [Section IV.D] an additional bias electric field  $F_b$ is introduced. The resulting electrostatic potential is described by  $W_b(z)=eF_bz$ 
for $z\in[-2d,2d]$, $W_b(z)=W_b(-2d)$ for $z<-2d$ and $W_b(z)=W_b(2d)$ for $z>2d$.

We apply the external magnetic field within the $xz$ plane, of the magnitude $B$ and the orientation defined by an angle $\alpha$ that it forms with the $z$ axis: ${\bf B}=(B_{x},0,B_{z})=(B\sin\alpha,0,B\cos\alpha)$.

\subsection{Time-dependent configuration interaction approach} 

In order to study the spin-valley dynamics of the confined carriers we {\it i)} calculate the single electron energy spectrum, 
{\it ii)} solve the Schr\"odinger equation for the few-electron states  with the Slater determinant basis built from the single-electron eigenstates, {\it iii)} solve the time-dependent Schr\"{o}dinger equation in the basis of the few electron eigenstates under the influence of the ac electric field. The results presented below provide an exact solution of the dynamics of the system 
composed of four electrons in states near the Fermi level.

In order to determine the single-electron states we use the atomistic tight-binding approach with the $p_z$ orbitals.  We solve the eigenproblem of  Hamiltonian
\begin{eqnarray}\label{ham1e}
&H_{1e}=&\sum\limits_{\{i,j,\sigma,\sigma'\}}(c_{i\sigma}^{\dagger}t_{ij}^{\sigma\sigma'}c_{j\sigma'}+h.c.)+ \\ &\sum\limits_{i,\sigma,\sigma'}c_{i\sigma}^{\dagger}&\left((W_{QD}({\bf r}_{i})+W_b({\bf r}_i))\delta_{\sigma\sigma'}+\frac{g_L\mu_{b}}{2}\boldsymbol{\sigma}^{\sigma\sigma'}\cdot{\bf B}\right)c_{i\sigma'}\nonumber.\label{ham}
\end{eqnarray}
The first sum in Eq. (\ref{ham1e}) accounts for the hopping between the nearest neighbor atoms. It runs over the $p_{z}$ spin-orbitals of the nearest
neighbor pairs of atoms, $c_{i\sigma}^{\dagger}\ensuremath{}(c_{i\sigma})$
is the particle creation (annihilation) operator at ion $i$ with
spin $\sigma$, and $t_{ij}^{\sigma\sigma'}$ is
the spin-dependent hopping parameter. The second sum accounts for  the external electric and magnetic fields, with $\delta_{\sigma\sigma'}$ standing for the Kronecker delta, $g_L=2$ for the Land\'e factor, $\mu_{b}$ the Bohr magneton and $\boldsymbol{\sigma}$ for the vector of Pauli matrices. 

In carbon nanotubes QDs \cite{kum,jesp,jar,chico}
the control over the spin is provided by mixing of the $\pi$ and $\sigma$ bonds that allows the carbon atomic spin-orbit coupling effects to appear in the electron band-structure \cite{review,klino,Md,Ando,St,kum}. 
	The spin-orbit interaction due to the curvature of the graphene plane introduces the spin dependence 
and spin mixing in the  hopping parameters $t_{ij}^{\sigma\sigma'}$. We apply the form of the parameters that accounts for both the folding of the graphene plane into the tube  \cite{Ando} 
and the curvature of the tube as a whole \cite{OsikaJPCM},

$t_{ij}^{\uparrow\uparrow}=(z_{i}|H|z_{j})+i\delta\cos\phi_{j}(z_{i}|H|x_{j})-i\delta\cos\phi_{i}(x_{i}|H|z_{j})+i\delta\sin\phi_{j}\sin\theta_{j}(z_{i}|H|y_{j})-i\delta\sin\phi_{i}\sin\theta_{i}(y_{i}|H|z_{j}),$

$t_{ij}^{\downarrow\downarrow}=(z_{i}|H|z_{j})-i\delta\cos\phi_{j}(z_{i}|H|x_{j})+i\delta\cos\phi_{i}(x_{i}|H|z_{j})-i\delta\sin\phi_{j}\sin\theta_{j}(z_{i}|H|y_{j})+i\delta\sin\phi_{i}\sin\theta_{i}(y_{i}|H|z_{j}),$

$t_{ij}^{\uparrow\downarrow}=-i\delta\sin\phi_{j}(z_{i}|H|x_{j})+i\delta\sin\phi_{i}(x_{i}|H|z_{j})-\delta(\sin^{2}\frac{\phi_{j}}{2}e^{i\theta_{j}}+\cos^{2}\frac{\phi_{j}}{2}e^{-i\theta_{j}})(z_{i}|H|y_{j})+\delta(\sin^{2}\frac{\phi_{i}}{2}e^{i\theta_{i}}+\cos^{2}\frac{\phi_{i}}{2}e^{-i\theta_{i}})(y_{i}|H|z_{j}),$

and

$t_{ij}^{\downarrow\uparrow}=-i\delta\sin\phi_{j}(z_{i}|H|x_{j})+i\delta\sin\phi_{i}(x_{i}|H|z_{j})+\delta(\sin^{2}\frac{\phi_{j}}{2}e^{-i\theta_{j}}+\cos^{2}\frac{\phi_{j}}{2}e^{i\theta_{j}})(z_{i}|H|y_{j})-\delta(\sin^{2}\frac{\phi_{i}}{2}e^{-i\theta_{i}}+\cos^{2}\frac{\phi_{i}}{2}e^{i\theta_{i}})(y_{i}|H|z_{j}),$\\

where $(\gamma_{i}|H|\gamma_{j})=V_{pp}^{\pi}{\bf n}(\gamma_{i})\cdot{\bf n}(\gamma_{j})+(V_{pp}^{\sigma}-V_{pp}^{\pi})\frac{({\bf n}(\gamma_{i})\cdot{\bf R}_{ji})({\bf n}(\gamma_{j})\cdot{\bf R}_{ji})}{{\bf |R}_{ji}|^{2}},$  $\gamma=x,\, y$ or $z$, $\gamma_{i}$ stands for the orbital 
of ion at ${\bf R}_{i}$ position and ${\bf n}(\gamma_{i})$ is a unit vector in the direction of orbital $\gamma_{i}$.
In the calculations we use the tight-binding Slater-Koster
parameters $V_{pp}^{\pi}=-2.66$ eV, $V_{pp}^{\sigma}=6.38$ eV \cite{Tomanek} and the spin-orbit coupling parameter $\delta=0.003$ \cite{Ando,Md}.
The interaction of the magnetic field with the orbital magnetic moments 
is taken into account by  the Peierls phase 
\[
t_{ij}^{\sigma\sigma'}(B)=t_{ij}^{\sigma\sigma'}(0)\exp(i\frac{2\pi}{\Phi_{0}}\int_{{\bf r}_{i}}^{{\bf r}_{j}}{\bf A}\cdot{\bf dl}),
\]
where $\Phi_{0}=h/e$ is the flux quantum, ${\bf B}=\nabla\times{\bf A}$,
and the Landau gauge ${\bf A}=(0,B_{z}x,B_{x}y)$ is applied.

\begin{figure}[htbp]
\includegraphics[width=0.8\linewidth]{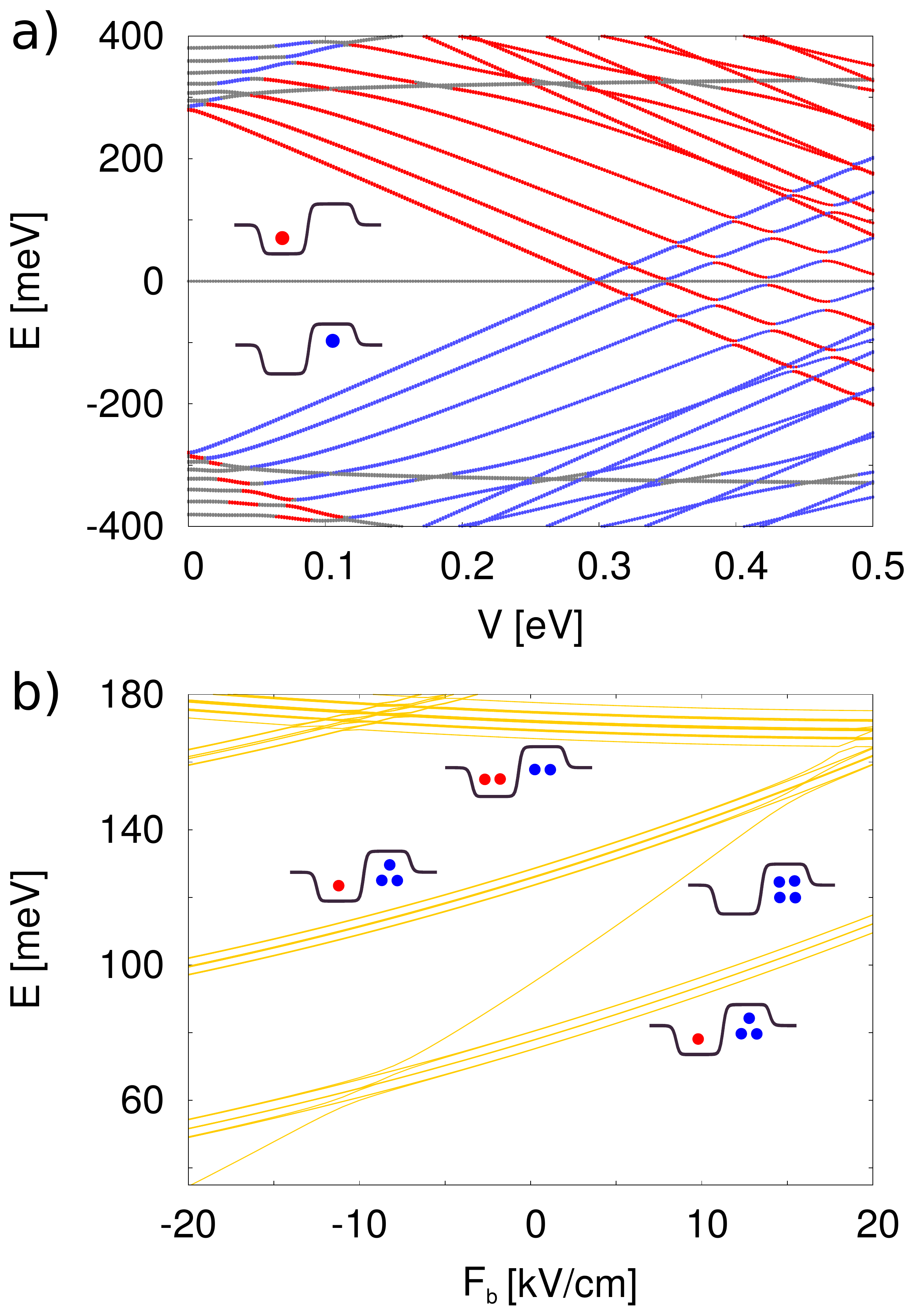}
\caption{ (a) Single electron energy levels as a function of the QD potentials $V=V_n=V_p$. Red/blue lines represent states with at least 40\% of the wavefunction localized within the n/p dot. (b) Four electron energy levels as a function of bias electric field $F_b$ for $V=0.21$ eV. The insets explain the QD charge occupation corresponding to each set of the energy levels.
} \label{widmoV}
\end{figure}

With the single-electron problem solved, we calculate the few electron eigenstates using the configuration interaction (CI) method. We are interested in (1e, 1h) charge configuration, in which 
in the $n$-type dot we have a single-electron (1e) in the conduction band and three electrons (or a single unoccupied state 1h) in the $p$-type dot (see Fig. \ref{schemat}).
Figure \ref{widmoV}(a) shows the single-electron energy spectrum as a function of the potential $V=V_n=V_p$. All the energy levels plotted in Fig. \ref{widmoV} are nearly fourfold degenerate with respect to the valley and spin.  
For the calculations we adopt $V=0.21$ eV 
and take into the basis several lowest-energy levels of the conduction band [the red curves in Fig. \ref{widmoV}(b)]  as well as the highest energy level of the valence band [the uppermost blue curve in Fig. \ref{widmoV}(a)].  We assume that all the energy levels below are fully occupied and do not participate in the dynamics of the system, which is governed
by the behavior of the last four electrons. 

The Hamiltonian for the interacting electron system  reads
\begin{equation}
H_{4e}=\sum_{a}\epsilon_{a}g_{a}^{\dagger}g_{a}+\frac{1}{2}\sum_{abcd}V_{ab;cd}g_{a}^{\dagger}g_{b}^{\dagger}g_{c}g_{d}, \label{4e}
\end{equation}
where $\epsilon_{a}$ is the energy of the $a$-th eigenstate of Hamiltonian $H_{1e}$ while $g_{a}^{\dagger}$ and $g_{a}$ are the creation and annihilation operators of the electron in the $a$-th state. The electron-electron interaction is taken into account in the second term of Eq. (\ref{4e}) with the matrix elements
\begin{eqnarray}
V_{ab;cd}&=&\langle\psi_{a}({\bf r_{1}},\boldsymbol{\sigma}_{1})\psi_{b}({\bf r_{2}},\boldsymbol{\sigma}_{2})|H_{c}|\psi_{c}({\bf r_{1}},\boldsymbol{\sigma}_{1})\psi_{d}({\bf r_{2}},\boldsymbol{\sigma}_{2})\rangle\nonumber \\
&=&\sum_{i,\sigma_{i};j,\sigma_{j};k,\sigma_{k};l,\sigma_{l}}\beta_{i,\sigma_{i}}^{a*}\beta_{j,\sigma_{j}}^{b*}\beta_{k,\sigma_{k}}^{c}\beta_{l,\sigma_{l}}^{d}\delta_{\sigma_{i};\sigma_{k}}\delta_{\sigma_{j};\sigma_{l}}\times \nonumber\\&&\langle p_{z}^{i}({\bf r}_{1})p_{z}^{j}({\bf r_{2}})|H_{C}|p_{z}^{k}({\bf r_{1}})p_{z}^{l}({\bf r_{2}})\rangle,
\end{eqnarray}
where $H_{c}$ is the electron-electron interaction potential
\[
H_{C}=\frac{e^{2}}{4\pi\epsilon\epsilon_{0}r_{12}}
\]
with $r_{12}=|\boldsymbol{r_{1}}-\boldsymbol{r_{2}}|$ and the dielectric
constant $\epsilon=9$ as for Al$_2$O$_3$ -- material which has been used as a substrate in the experimental setups \cite{lairdpei}.
The coefficients $\beta_{i,\sigma_{i}}^{a}$ define contributions of $p_{z}^{i}$
orbitals of spin $\sigma_{i}$ to the single-electron eigenstate $a$.
In the calculations we use the two-center approximation \cite{tc}.
The on-site integral ($i=j$) we approximate by $\langle p_{z}^{i}p_{z}^{j}|\frac{e^{2}}{4\pi\epsilon_{0}r_{ij}}|p_{z}^{i}p_{z}^{j}\rangle=16.522$
eV \cite{potasz} and for $i\neq j$ we use the formula
$\langle p_{z}^{i}p_{z}^{j}|\frac{1}{r_{ij}}|p_{z}^{k}p_{z}^{l}\rangle=\frac{1}{r_{ij}}\delta_{ik}\delta_{jl}$ \cite{Osambi}.
The atomistic approach used here accounts for all intervalley effects \cite{rontani} that accompany the short range component of the Coulomb potential.
Moreover, the present approach is not limited by the low-energy continuum approximation, and covers an ample variation of the external potential necessary for formation of an ambipolar quantum dot within the tube.

The energy spectrum is given in Fig. \ref{widmoV}(b) as a function of the bias field. The slope of the lines is determined by the electric dipole moment of the system  
i.e., the electron distribution between the dots. The ground-state corresponds to the (1e,1h) charge configuration that is focused below. The nonzero bias field is considered
in Section IV.D, elsewhere we take $F_b=0$.

We simulate the valley and spin transitions driven by external ac field by solving the time-dependent Schr\"{o}dinger equation with Hamiltonian
\begin{equation}
H'(t)=H_{4e} + \sum_{j=1}^4 eF_0 z_j \sin(\omega t),
\end{equation}
where $F_0$ is the ac electric field amplitude and $\omega$ is its  frequency. Using the eigenstates $\Psi_n$ of  Hamiltonian $H_{4e}$ we construct  basis in which the time-dependent Schr\"{o}dinger equation is solved
\begin{equation}
\Psi({\bf r}_{1...4}, {\boldsymbol \sigma_{1...4}}, t) =  \sum_n c_n(t) \Psi_n({\bf r}_{1...4}, {\boldsymbol \sigma_{1...4}}, t) e^{-\frac{iE_nt}{\hbar}}.
\end{equation}
In this basis the Schr\"{o}dinger equation $i\hbar\frac{\partial\Psi}{\partial t}=H'\Psi$ takes the form
\begin{equation}
i\hbar \dot{c}_k(t)=\sum_n c_n(t) eF_0 \sin(\omega t) \langle\Psi_k|z|\Psi_n\rangle e^{-\frac{i(E_n-E_k)t}{\hbar}}. \label{cndot}
\end{equation}
We discretize the time in equation (\ref{cndot}) and calculate the coefficients $c_k(t)$ using the Crank-Nicolson algorithm. 

\subsection{Floquet approach}
The direct solution of the   time-dependent  Schr\"odinger method is supported with the Floquet theory.
We use the Floquet Hamiltonian method \cite{Shirley, Chu} to describe the dynamics of the system. For Hamiltonian $H'(t)$ that is periodic in time with the period $T=2\pi/\omega$, the Floquet theorem asserts the existence of the solution of the Schr\"{o}dinger equation 
\begin{equation}
 i\hbar\frac{\partial\Psi}{\partial t}=H'\Psi
\end{equation}
of the form
\begin{equation}
\Phi(t)=\phi(t) e^{-i\epsilon t/\hbar},
\end{equation}
where $\phi(t)$ is periodic in time. The $\Phi$ functions are called quasienergy eigenstates (QES) and upon the Fourier series expansion can be expressed by
\begin{equation}
\Phi_\alpha=e^{-i\epsilon_\alpha t/\hbar}\sum_{n=-\infty}^{\infty}\sum_{\beta}v_{\alpha\beta}^{(n)} e^{-in\omega t} |\beta\rangle ,
\end{equation}
where $\alpha$ enumerates the quasi-energy state, $\epsilon_\alpha$ is the quasienergy, $\beta$ are eigenfunctions of the Hamiltonian $H_{4e}$ and $v_{\alpha\beta}^{(n)}$ are coefficients expanding the quasi-energy state in the basis of $\beta$ eigenstates.
Since functions $\Phi_\alpha$ satisfy the equation $(H'-i\hbar\frac{\partial}{\partial t})\Phi_\alpha=0$, after expanding it in the basis of $\beta$ eigenstates we obtain 
\begin{equation}
\sum_n \sum_{\beta} [ \langle\alpha|H'^{(m-n)}|\beta\rangle - (\epsilon_\alpha+m\omega)\delta_{mn}\delta_{\alpha\beta} ]v_{\alpha\beta}^{(n)} =0,
\end{equation}
where
\begin{equation}
H'^{(n)}=\frac{1}{T}\int_0^T H'(t)e^{in\omega t} dt .
\end{equation}
For the case of the perturbation $eF_0 z \sin(\omega t)$ the matrix elements $\langle\alpha|H'^{(m-n)}|\beta\rangle$ read
\begin{eqnarray}
\langle\alpha|H'^{(m-n)}|\beta\rangle &=&E_\alpha\delta_{\alpha\beta}\delta_{mn}+ 
\frac{i}{2}\delta_{m-n,1}\langle\alpha|eF_0 z|\beta\rangle + \nonumber \\
&-&\frac{i}{2}\delta_{m-n,-1}\langle\alpha|eF_0 z|\beta\rangle .
\end{eqnarray}
Finally, one writes the time-independent Floquet Hamiltonian $H_F$ with the matrix elements defined by
\begin{equation}
\langle\alpha m|H_F|\beta n\rangle=\langle\alpha|H'^{(m-n)}|\beta\rangle+n\hbar\omega\delta_{\alpha\beta}\delta_{mn}
\end{equation}
in a block form we can express the matrix of the Hamiltonian $H_F$ as
\[
\begin{bmatrix}
\ddots & \cdots & \cdots & \cdots & \cdots  & \cdots & \reflectbox{$\ddots$} \\
\vdots & A+2\hbar\omega I & B & 0 & 0 & 0  & \vdots \\
\vdots & -B & A+\hbar\omega I & B & 0 & 0  & \vdots \\
\vdots & 0 & -B & A & B & 0 & \vdots \\
\vdots & 0 & 0 & -B & A-\hbar\omega I & B  & \vdots \\
\vdots & 0 & 0 & 0 & -B & A-2\hbar\omega I   & \vdots \\
\reflectbox{$\ddots$} & \cdots & \cdots & \cdots & \cdots & \cdots & \ddots
\end{bmatrix}
\]
with matrix $A$ containing on the diagonal the energies of the eigenstates $|\alpha\rangle$ and matrix $B$ built of the elements $\frac{i}{2}\langle\alpha|eF_0 z|\beta\rangle$. 

We solve the eigenproblem of the Hamiltonian $H_F$ and obtain a set of eigenenergies $\epsilon_l$ and eigenstates $|\epsilon_l\rangle$. Using the result we can calculate the time averaged (average over initial time $t_0$ and ac pulse duration $t-t_0$) probability of the transition between $\alpha$ and $\beta$ states 
\begin{equation}
P_{\alpha\rightarrow \beta}=\sum_n \sum_l |\langle \beta n | \epsilon_l \rangle \langle \epsilon_l|\alpha 0 \rangle |^2 .
\end{equation}
In order to obtain convergent results for the probabilities $P_{\alpha\rightarrow \beta}$ a sufficient number of the harmonics $\pm n\hbar\omega$ must be included into the $H_F$ matrix.

\begin{figure}[htbp]
\includegraphics[width=0.7\linewidth]{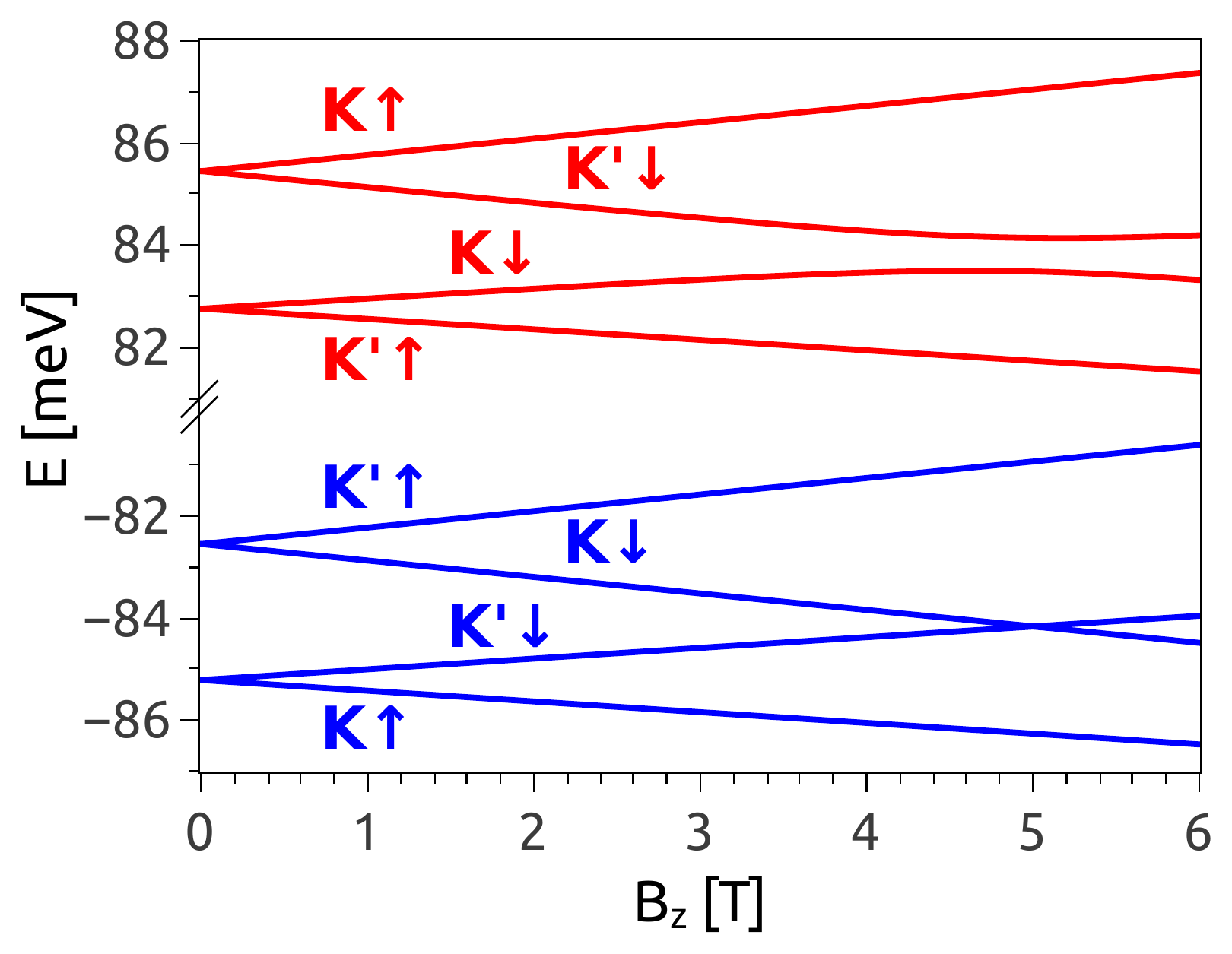}
\caption{ Energy of the dot-localized single electron states of the top of the valence (blue lines) band and the bottom of the conduction band (red lines) as a function of $B_z$. 
} \label{widmoB1el}
\end{figure}

\begin{figure}[htbp]
\includegraphics[width=0.95\linewidth]{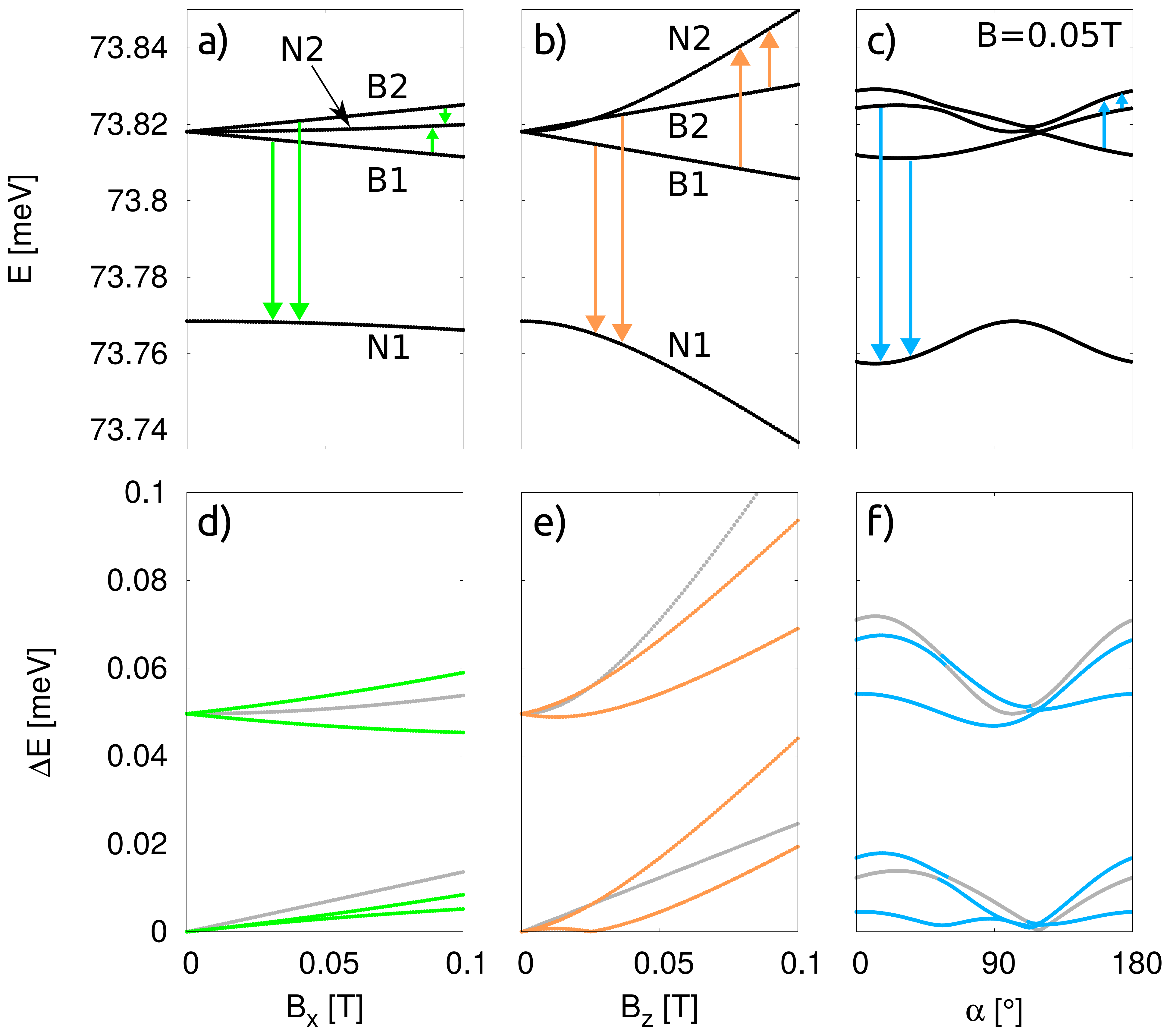}
\caption{ (a-c) Energy of the four-electron lowest-energy states as a function of $B_x$, $B_z$ and $\alpha$, respectively. The color arrows represent the transitions between Pauli blocked and nonblocked states. (d-f) Transition lines corresponding to the arrows on the plots (a-c).
} \label{widmaB}
\end{figure}

\begin{figure}[tb!]
\includegraphics[width=0.8\columnwidth]{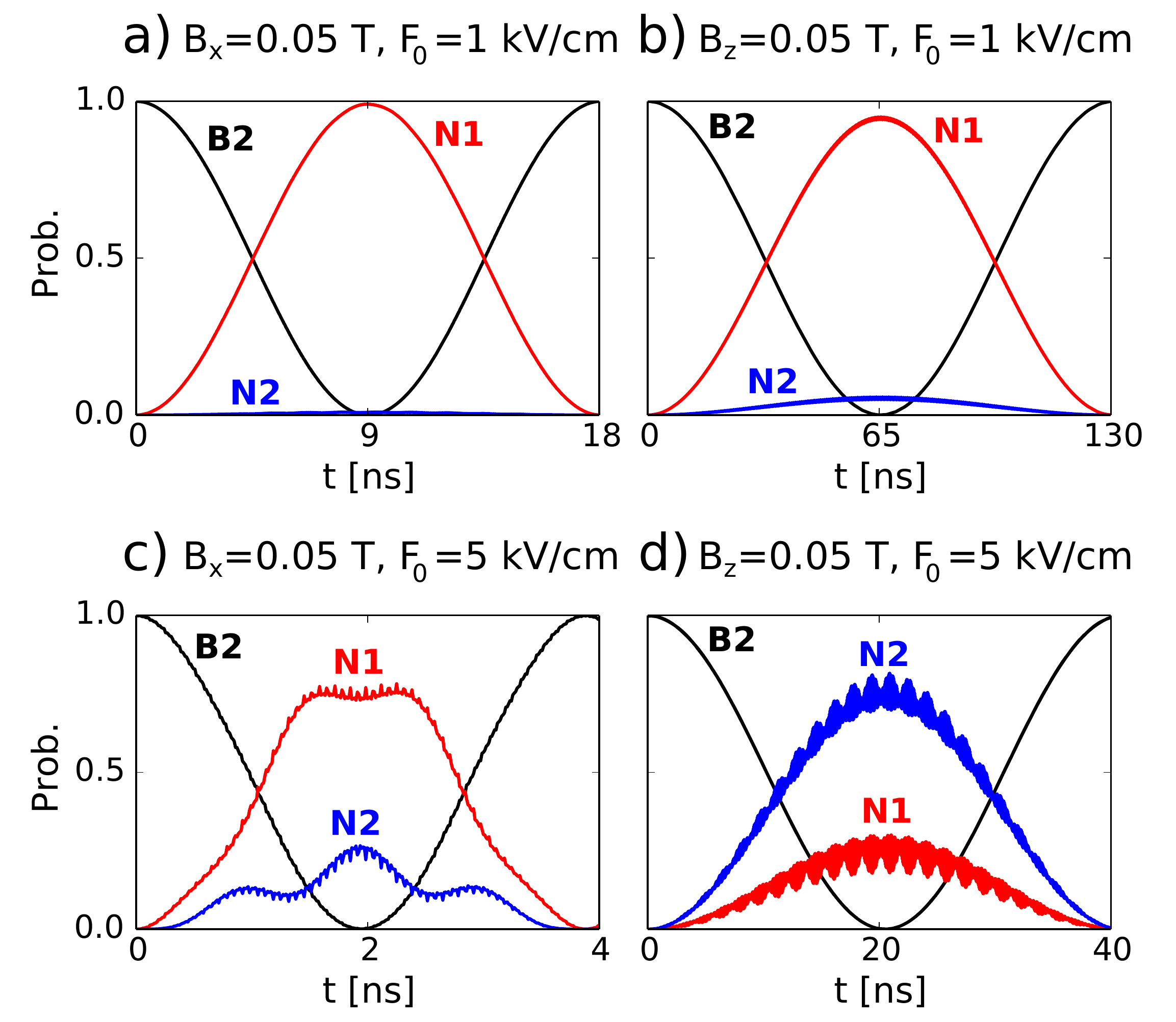}
  \caption{ N1, N2 and B2 occupation probabilities during the B2$\rightarrow$N1 resonant transition at four different magnetic fields and amplitudes of the ac electric field. 
  } \label{czas}
\end{figure}

\begin{figure*}[htbp]
\includegraphics[width=0.99\linewidth]{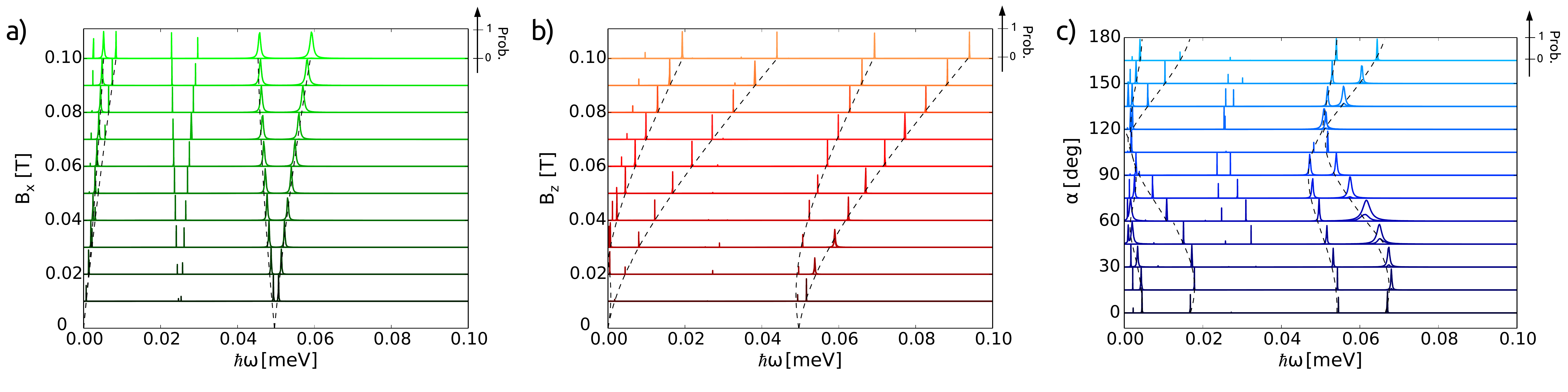}
\caption{Maximal probability of the transition between Pauli blocked and non-blocked states (during 200 ns time evolution) as a function of the ac field frequency $\omega$ and (a) $B_x$, (b) $B_z$ and (c) angle $\alpha$ for $B=0.05$ T. The dashed lines indicate the $B\rightarrow N$ transition energies as calculated from the energy spectrum. 
} \label{skany}
\end{figure*}


\begin{table}
\begin{tabular}{l|l} 
(1e,1h) state & occupied single-electron states  \\  \hline 
N1 & pK$\uparrow$  pK'$\downarrow$  pK$\downarrow$  nK'$\uparrow$ \\
N2 & pK$\uparrow$  pK'$\downarrow$  pK'$\uparrow$  nK$\downarrow$ \\
B1 & pK$\uparrow$  pK'$\downarrow$  pK$\downarrow$  nK$\downarrow$ \\
B2 & pK$\uparrow$  pK'$\downarrow$  pK'$\uparrow$  nK'$\uparrow$ \\
\end{tabular}
\caption{The dominant occupation of the single-electron energy levels of Fig. \ref{widmoB1el} in the lowest-energy (1e,1h) states of Fig. \ref{widmaB}(a,b,c) for nonzero magnetic field ($B=0.1$ T). The first column indicates the energy level of Fig. \ref{widmaB}(a,b,c), and the remaining columns the energy levels occupied in the $p$ and $n$ type dot.
 } \label{tab}
\end{table}

\begin{table}
\begin{tabular}{|c||c|c|c|c|c|}
\hline
& N1 & B1 & B2 & N2 & (0e,0h) \\ \hline \hline
N1 & 15.84 &  9.53$\cdot 10^{-5}$ & 3.23$\cdot 10^{-4}$ & 2.12$\cdot 10^{-2}$ & 1.03 \\ \hline
B1 & 9.53$\cdot 10^{-5}$ & 15.79 & 1.62$\cdot 10^{-6}$ & 8.22$\cdot 10^{-5}$ & 2.94$\cdot 10^{-3}$ \\ \hline
B2 &  3.23$\cdot 10^{-4}$ & 1.62$\cdot 10^{-6}$ & 15.79 & 2.51$\cdot 10^{-4}$ & 7.31$\cdot 10^{-3}$ \\\hline
N2 & 2.12$\cdot 10^{-2}$ &8.22$\cdot 10^{-5}$& 2.94$\cdot 10^{-3}$& 15.80 & 0.433 \\\hline
(0e,0h) &1.03&2.94$\cdot 10^{-3}$&2.51$\cdot 10^{-4}$ &0.433 & 33.12 \\\hline
\end{tabular}
\caption {Dipole matrix elements $\langle \psi_i|z|\psi_f \rangle$  between the energy levels marked in Fig. 5(b) by B1,B2,N1 and N2 for (1e,1h) charge configuration and the ''vacuum state'' (0e,0h) with no electrons in the $n$-dot and fully filled $p$-dot (4 electrons in the dot) for $B_z=0.05$ T.
The dominant Slater determinants for (1e,1h) states are indicated in Table I. 
The vacuum state  (0e,0h) is the non-degenerate energy level that grows fastest with the bias field $F_{b}$ in Fig. 3(b). The results are given in nanometers.  }
\end{table}

\section{Spectra in external magnetic field}

The single-carrier energy levels confined in the $n$- and $p$-type quantum dots are displayed in Fig. \ref{widmoB1el} and the calculated lowest energy levels of the (1e,1h) charge configuration in Fig. \ref{widmaB}(a,b,c). 
The single-electron states occupied in the dominant configurations for the (1e,1h) states at $B=0.1$ T are listed in Table I. 

In the four lowest-energy (1e,1h) states of Fig. \ref{widmaB}(a-c) -- in the dominant configurations -- two of the three electrons of the $p$ dot occupy the two-lowest energy states of the valence band (p$K\uparrow$ and p$K'\downarrow$ in Fig. \ref{widmoB1el}).  
The third electron of the $p$ dot occupies one
of the two-highest energy levels of the valence band (p$K\downarrow$ or p$K'\uparrow$). Finally, the single electron in the $n$ type dot 
occupies one of the two-lowest energy states  of the conduction band (n$K\downarrow$ or n$K'\uparrow$).
The low-energy spectrum at $B=0$ [Fig. \ref{widmaB}(a,b)] consists of a ground-state singlet and an excited state triplet. 
In the states denoted by B1 and B2 in Fig. \ref{widmaB} the last two electrons are spin-valley polarized, i.e. in B1 the last electron in both $n$ and $p$ dot
occupies the $K\uparrow$ energy level, and for B2 the occupied spin-valley level is $K'\uparrow$. Both these states are blocked, i.e.  in terms of the {\it dominant} spin-valley configurations of  Table I,  the electron of the $n$ dot
is forbidden to pass to the $p$ dot by the Pauli exclusion principle, since the state
with the same spin and valley is occupied in the $p$ dot.
This is not the case for the other two states -- denoted by N1 and N2 in Fig. \ref{widmaB}. These states  are referred to as ''nonblocked'' in the following. The nonblocked states enter into an avoided crossing that is opened at $B=0$ by the exchange integral \cite{Osambi,eo_bent}, hence their non-linear dependence on $B$ near 0T.  

The orbital magnetic dipole moment due to the electron circulation around the tube \cite{orbital} is oriented parallel or antiparallel to the axis of the tube \cite{review}.
Moreover, the circumferential spin-orbit interaction fixes the projection of the electron spin on the orbital magnetic moment \cite{kum}. Thus, the spins of the lowest-energy states are nearly polarized along the axis of the tube and
 a weak magnetic field applied in the $x$ direction does not affect the energy levels of electrons confined within  a straight CNT.
The entire dependence of the energy levels on ${\bf B}=(B_x,0,0)$ that is visible in Fig. \ref{widmaB}(a) results from  the bent of the CNT. For the considered radii of the bent [$R_n$, $R_p$ in Fig. 2] the reaction of the energy levels on the magnetic field ${\bf B}=(0,0,B_z)$ is still much stronger 
 [Fig. \ref{widmaB}(b)]. The reaction anisotropy to the magnetic field vector leads to the distinct dependence of the spectrum on the angle $\alpha$ that the vector ${\bf B}$ forms with the $z$ axis [Fig. \ref{widmaB}(c)].

\section{Spin-valley transitions}

\subsection{Weak ac field} 

The lifting of the Pauli blockade in states B1 and B2 is achieved via transitions to one of the states N1 or N2 \cite{pei,lairdpei,eo_bent,li}  -- see the arrows in Fig. \ref{widmaB}(a-c) -- that is induced by the ac electric field. Figures \ref{widmaB}(d-f) show the energy difference
between the blocked  and nonblocked B-N levels (color lines). For completeness, the energy difference between the blocked pair of energy levels B1-B2 (the gray line that starts at zero energy at $B=0$) and unblocked N1-N2 levels (the higher energy gray line) were also plotted.

Figure \ref{czas} shows the time-resolved occupation probabilities obtained for 
the driving  frequency set for the B2$\rightarrow$N1 resonant transition with the electron
initially in state B2. 
 For a weak ac field amplitude of 1 kV/cm [Fig. \ref{czas}(a,b)] the transition is much faster for 
the magnetic field that is oriented perpendicular to the axis of the CNT ${\bf B}=(B_x,0,0)$
than for $(0,0,B_z)$. The transition from B2 to N1 involves both the valley and the spin inversion in pK'$\uparrow\rightarrow$ pK$\downarrow$ [Table I]. For a straight CNT spins are polarized in the $z$ or $-z$ directions. 
For $B$ oriented along the $x$ direction the energy change is weaker [Fig. \ref{widmaB}(a)] but this
field orientation  contributes in mixing the $\sigma_z$ eigenstates and to opening the channel for the spin inversion, hence the shorter
transition times.

Figure \ref{skany} shows the maximal transition probability from B1 and B2 states to the nonblocked ones N1 and N2 as functions of the 
driving ac frequency for a number of magnetic field values [Fig. \ref{skany}(a,b)] and orientation angles [Fig. \ref{skany}(c)] for $F_0=1$ kV/cm -- i.e. the one considered in Fig. \ref{czas}(a,b).  For each value of the magnetic field in Fig. \ref{skany} we plot two curves for the initial state set at either B2 and B1. The dashed lines 
indicate the nominal transition energies which agree very well with the peaks position.
We can see that the width of the lines for $B_x$ is larger which is consistent with the shorter transition times [Fig. \ref{czas}(a,b)]
for the two-level Rabi transitions. For this --  weak oscillations amplitude -- considered in Fig. \ref{skany} (a) the $g$ factors for the transitions can
be exactly estimated from the energy spectra. 

Let us briefly comment on the relation of the energy differences between the blocked
and nonblocked energy levels and the experimental EDSR transitions (Fig. 2 of Ref. \cite{lairdpei}):
{(\it i)} The dependence over the orientation angle of the magnetic field in the $(x,z)$ plane 
of both the experimental (Fig.2(c) of Ref. \cite{lairdpei}) and the present (Fig. 5(f)) results  exhibits a pair of lines near zero transition energy/frequency
separated by a gap from another pair of levels. The lines change in phase with the maxima near 0 and 180$^\circ$
and a minimum near 0. {\it (ii)} In the upper pair of lines  as a function of $B_x$ [Fig. 2(a) of Ref. \cite{lairdpei}] 
one of the lines increases and the other decreases with the magnetic field. The effective $g$ factor  extracted from the slope for the increasing line 
$g=\frac{1}{\mu_B} \frac{dE}{dB}$ of the experimental paper is 1.8 while for the data of Fig. 5 the value is 1.67 (we take the derivative near 0.05 T). 
For the decreasing line we find -0.68. Although no $g$ factor estimate has been given for that line in the experiment, from Fig.4(h) in Ref. \cite{li} we can assert that the slope of the decreasing line is less steep than for the increasing one, thus $|g|<1.8$. For the lower lines we find $g$ factors of 1.49 and 0.86, while the experimental data provides 1.9 for one of the lines. For nonzero $B$ only a single lower line is resolved in the experiment. {\it (iii)} The largest deviation for the $g$ factors is found for $B_z$  orientation of the field: we find 8.67 and 4.42 for the $g$ factors in the upper branch, while the experimental paper produces 4.5 and 3 respectively. A possible explanation for the deviation of the actual transition spectrum from the spacing derived from the energy spectrum is provided below.


\begin{figure}[htbp]
\includegraphics[width=0.95\linewidth]{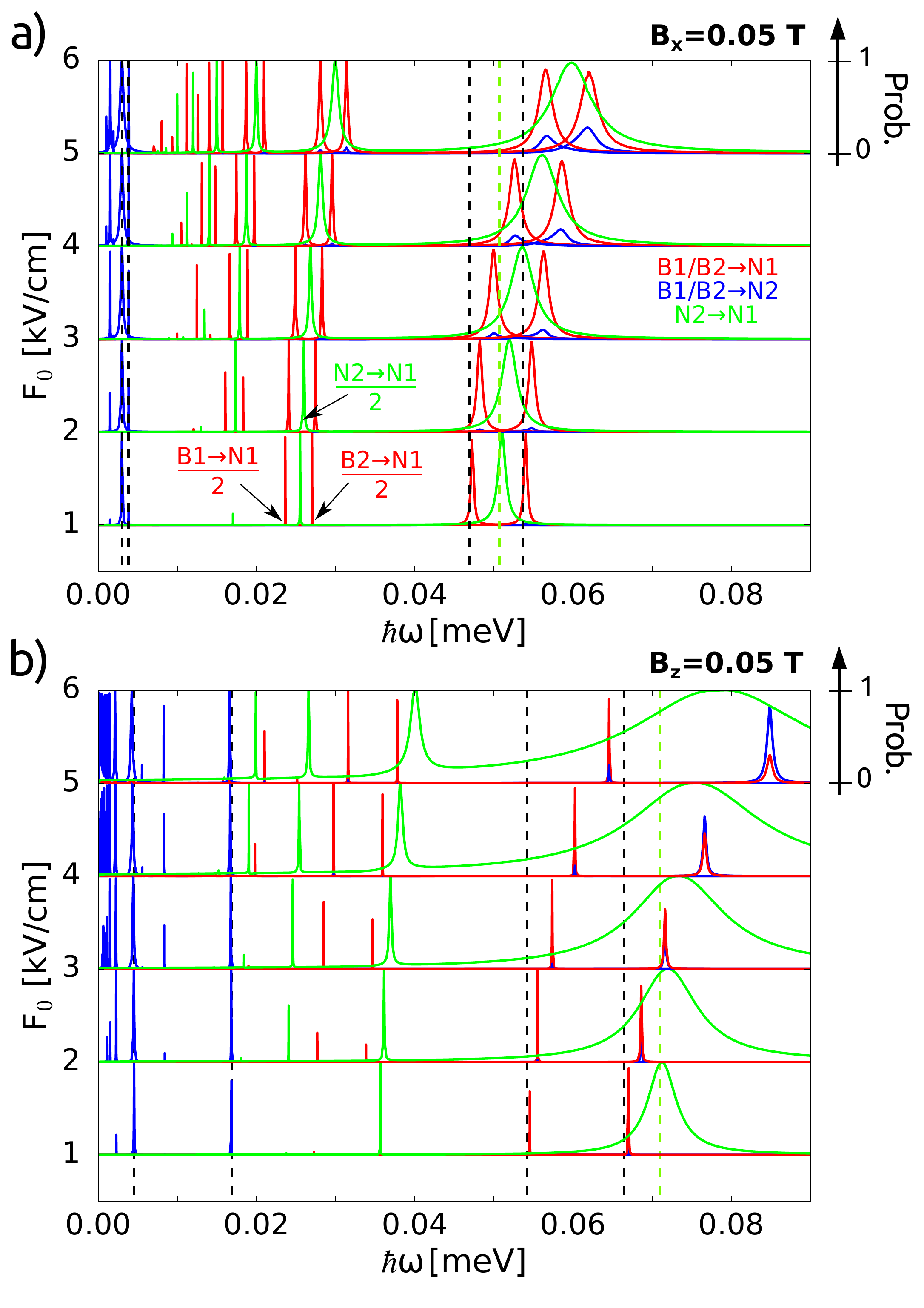} 
\caption{Maximal occupation probability obtained for a solution of the Schr\"odinger equation for 200 ns as a function of the driving field frequency and
amplitude for (a) $B_x=0.05$ T and (b) $B_z=0.05$ T.
The plots show the simulations for both B1, B2 set in the initial condition that evolve to N1 (red line) or N2 (blue line) states. The green line indicates the maximal occupation probability of N1 state for N2 in the initial condition. 
The dashed vertical line show the energy differences for B$\rightarrow$ N transition (black lines) and N2$\rightarrow$N1 transition (green line).
} \label{nunu0}
\end{figure}

\begin{figure}[htbp]
\includegraphics[width=0.9\linewidth]{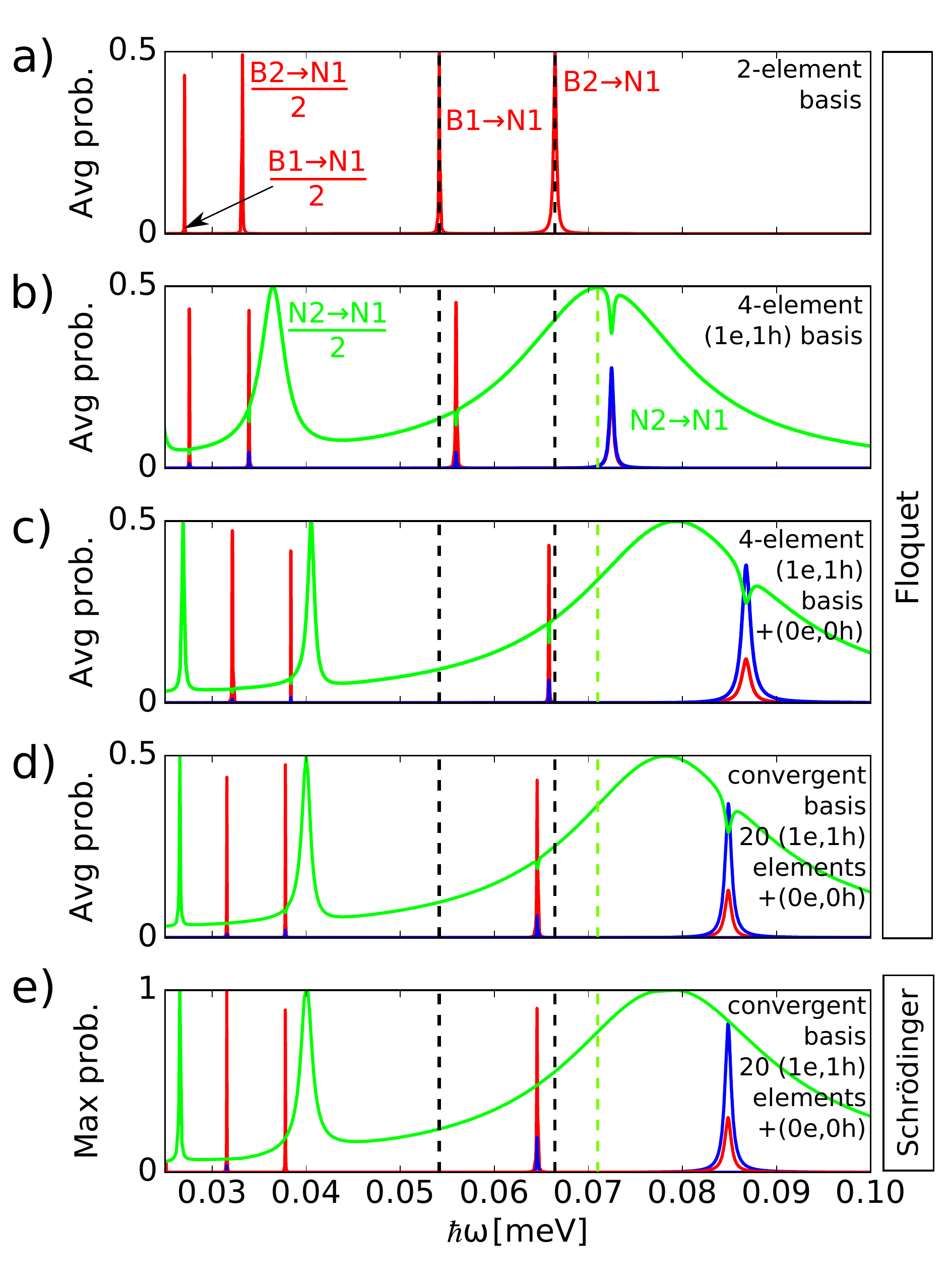} 
\caption{The transition spectra for $B_z=0.05$ T and $F_0=5$ kV/cm for varied basis of few-electron stationary Hamiltonian eigenstates. } \label{nunubasis}
\end{figure}

\subsection{Strongly driven system}

We analyze an effect of an amplification of the ac driving field on the dynamics of the system. In Fig. \ref{nunu0} we plot the transition spectra for amplitudes $F_0=1,2,...,5$ kV/cm at two different magnetic field orientations -- $B_x=0.05$ T and $B_z=0.05$ T.
In both cases the increase of $F_0$ yields a few interesting effects: (i) broadening of the resonant peaks, (ii) shifts of the resonant transition energies (especially large for the upper branch lines), (iii) emergence of the B2 $\rightarrow$ N2 transition at B2 $\rightarrow$ N1 resonant peaks, and (iv) appearance of fractional resonances at the fractions of the resonant frequencies. 

Widening of the resonant peaks is a signature of the acceleration of the transitions. For the case of Fig. \ref{czas}(c,d) an increase of the ac electric field amplitude $F_0$ from 1 to 5 kV/cm shortens the transition time slightly less than 5 times. 
The ac frequency in Fig. \ref{czas}(c,d) is targeted to B2$\rightarrow$N1 transition.
Nevertheless, one observes also an appearance of N2 energy level, although the B2$\rightarrow$N2 transition is strongly off resonance with the driving frequency. 
The energy for the direct B2$\rightarrow$N2 transition is 10 times smaller then to N1, and the dipole element for the transitions to N1 and N2 are similar [see Table II]. 

Note, that in Table II the dipole matrix elements between B and N states are non-zero only because of the small admixtures of the opposite spin and valley which appear in the single-electron states -- indicated in Table I  -- due to the presence of the intervalley and spin-orbit coupling. The diagonal elements have an interpretation of the dipole moment of the state.  For the quadruple of (1e,1h) states the dipole moment is  similar, and for (0e,0h) it is larger which agrees with the slope of the energy levels in Fig. \ref{widmoV}(b) with $F_b$.

Better insight into origin of the resonant frequency shifts can be provided by the analysis of the convergence of the basis.
This can be useful also for the discussion of the appearance of N2 in the dynamics at the frequency
which is set to transition to N1.  Figure \ref{nunubasis}
presents the average occupation probability for $F_0=5$ kV/cm and $B_z=0.05$ T as obtained
from the Floquet theory and a growing number of basis elements. We can see that the transition
spectrum gets blue-shifted from the two-level Rabi transition [Fig. \ref{nunubasis}(a)] with the inclusion
of the entire quadruple (B1,B2,N1,N2) of the lowest-energy (1e,1h) levels [Fig. \ref{nunubasis}(b)]. 
Moreover formation of the two-photon resonance for N1$\rightarrow$N2 transition is observed at half the 
frequency for the direct resonance.
Inclusion of the vacuum state (0e,0h) (empty $n$ dot, four electrons filling completely the $p$ dot energy levels) as the fifth element to the basis provides even stronger blueshift and produces  a spectrum nearly identical with the  convergent one. 
The transitions are blue-shifted as in the Bloch-Siegert shift \cite{blochs,Shirley}, however the effects
of Fig. \ref{nunu0} do not follow the dependence on $F_0$ for the Bloch-Siegert transitions
since they involve more than two energy levels. 
Note, that for $B_z=0.05$ T the order of the transition lines is changed with respect to the energy spectrum (dashed vertical lines): the N1$\rightarrow$N2 transition enters between the B$\rightarrow$N1 lines. 
 
We can see that the electron tunneling from (1e,1h) quadruple to the nondegenerate (0e,0h) state has a pronounced
effect on the transition spectrum. 
Analyzing the right hand side of Eq. (7) we found (see Supplement \cite{supplement}) that for the transitions of Fig. 6(c,d) (an exact basis and a resonant frequency was used) the electron from the initial state B2  is transferred most effectively to the vacuum state (0e,0h) -- for which the transition matrix element [Table II] is the largest (the transition rate B2$\rightarrow$(0e,0h) is larger by an order of magnitude than for B2$\rightarrow$N transitions).  Note, that this transition is off resonance, since the ac frequency in Fig. 6
is tuned to B2$\rightarrow$N1 transition and the (0e,0h) state is for $F_b=0$ about 20 meV higher in the energy [Fig. 3(b)] than the (1e,1h) ground-state.
Note, that for the AC field with amplitude $F_0=5$ kV/cm the energy difference between the lowest-energy (1e,1h) state and the (0e,0h) state is quite large 
within the driving period and 
varies  between 10 meV  and 30 meV  -- see Fig. 3(b) for $F_b\in (-5,5)$ kV/cm.

Beside the resonant B2$\rightarrow$N1 transition, in Fig. 6(c,d)  both N1 and N2 states get occupied by transitions from the vacuum state (0e,0h)
-- these states are more strongly coupled to the vacuum state (0e,0h) than to one other [Table II]. The vacuum state serves
as an intermediate one for transitions to N1 and N2 states.  The transition (0e,0h)$\rightarrow$N1/N2
is immediate which prevents accumulation of the electron in the vacuum  state  (0e,0h). The (0e,0h) occupation probability in the conditions of Fig. 6(d)  is 2\textperthousand\;  at most [see Fig. S1 in the Supplement \cite{supplement}]. 
The transition N2$\leftrightarrow$N1 is nearly in resonance with the driving frequency resonant for the B2$\rightarrow$N1 transition, so for the conditions of Fig. \ref{czas}(d) we find that N2$\leftrightarrow$N1 transitions are by a factor of 1.5 to 2 more effective than for the N$\leftrightarrow$(0e,0h) ones [see Fig. S2 in the Supplement \cite{supplement}].
The N2 state has also been observed in Fig. 6(b) for the smaller amplitude of $F_0=1$ kV/cm. For this amplitude the vacuum state (0e,0h) does participate in the transition with adequately lower maximal occupation probability of about 0.05\textperthousand\;.

\begin{figure}[htbp]
\includegraphics[width=0.95\linewidth]{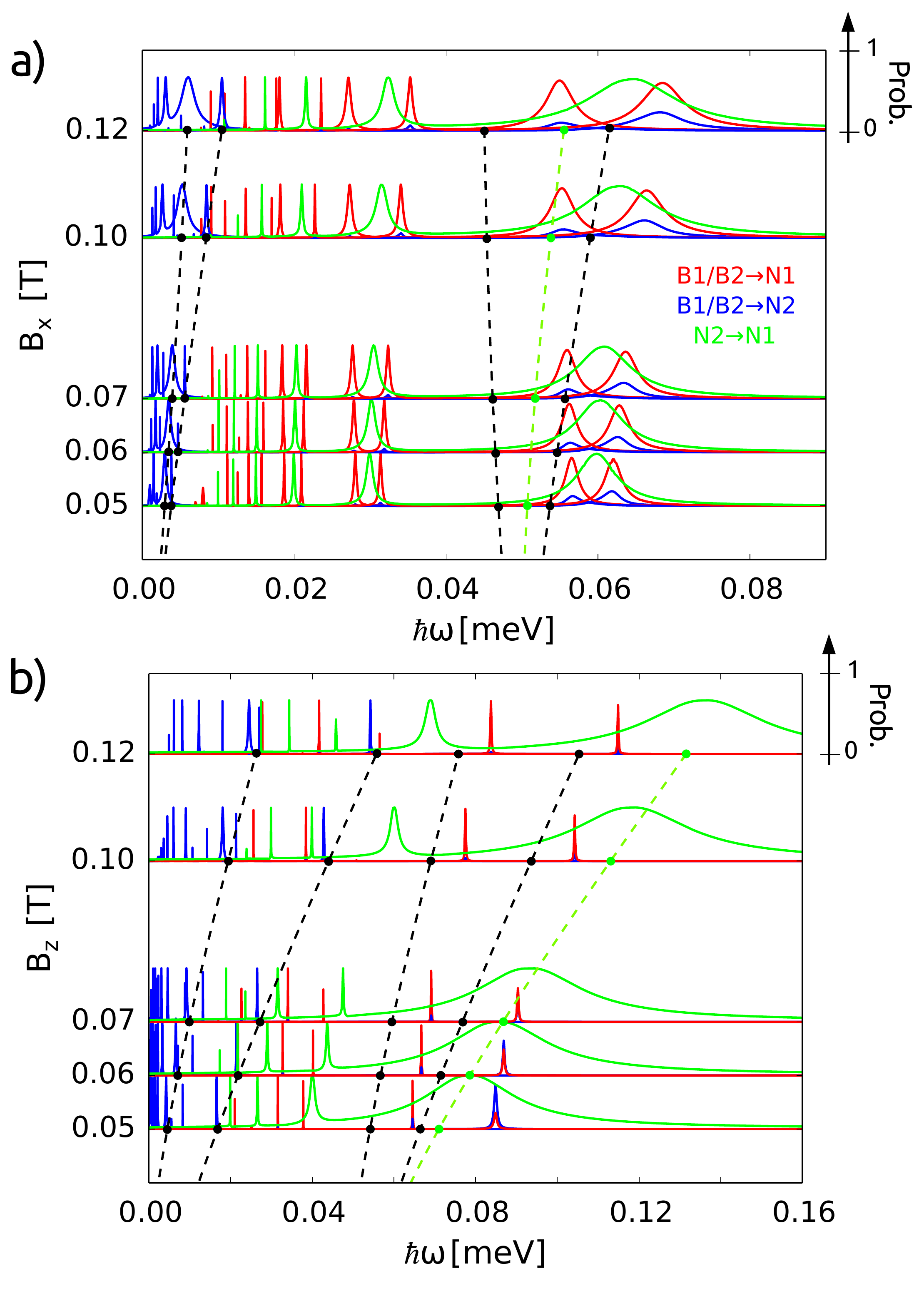} 
\caption{ Maximal occupation probability obtained for a solution of the Schr\"odinger equation for 200 ns as a function of the driving field frequency and
the magnitude of (a) $B_x$ and (b) $B_z$ magnetic field. The ac driving field amplitude was set to 5 kV/cm.
} \label{nunub}
\end{figure}

\subsection{The effective $g$ factors for transitions lifting the spin-valley blockade} 
The dependence of the transition spectra on the magnetic field is given in Fig. \ref{nunub}. Let us focus on the direct B1/2$\rightarrow$N1 and N2$\rightarrow$N1 transitions -- the ones
of  the high energy branch of the plots. For $B_x$ orientation of the field -- the three maxima 
are blue-shifted with respect to the energy difference, but the shift does not strongly depend on the magnetic field. 
Hence, the effective $g$ factors for the $B_x$ orientation are similar to the ones obtained from the energy spectra -- see the upper half of Table III.  A slight reduction of the absolute value of the $g$ factors is observed in the transition lines for  $F_0\geq 3$ kV/cm.

For the magnetic field oriented along the $z$ axis the $g$ factors deviate more significantly from the ones obtained from the energy spectra. In the energy spectra for $B_z$ field the N2 level shifts promptly up the energy scale off B2 and B1 energy levels, and the transition N2$\leftrightarrow$N1 energy separates from B$\rightarrow$N1 energy -- on contrary to what is observed for $B_x$ field. When N2$\leftrightarrow$N1 shifts off the B1/2$\rightarrow$N1 transition energies, the blueshift of their energies is reduced, hence the reduction for the $g$ factors (lower part of Table III). The $g$ factors -- as taken from the spectra
were by a factor of 1.5 to 2 larger than in the experiments. The $g$ factors as calculated from the transition spectra are significantly decreased for larger $F_0$. Moreover, the values for both B$\rightarrow$N1 transitions  
produce similar $g$ factors at larger $F_0$ -- while from the energy spectra
one was nearly 2 times larger then the other.

\begin{table}
 \begin{tabular}{l|l l} 
& $g$ factors  & for $B_x$  \\  &  B2$\rightarrow$N1 & B1$\rightarrow$N1 \\\hline \hline 
$E$ spectrum & 1.7 & -0.7 \\ \hline 
transitions for $F_0=3$ kV / cm& 1.6 & -0.6 \\ \hline 
transitions for $F_0=5$ kV / cm&  1.3 & -0.5 \\ \hline  
experiment \cite{lairdpei} & 1.8 & - \\ \hline \hline 
& $g$ factors  & for $B_z$  \\  &  B2$\rightarrow$N1 & B1$\rightarrow$N1 \\\hline \hline 
$E$ spectrum & 8.7 & 4.4 \\ \hline 
transition lines for $F_0=3$ kV / cm& 7.4 & 4.2 \\ \hline 
transition lines for $F_0=5$ kV / cm&  3.5 & 3.7 \\ \hline  
experiment \cite{lairdpei}  & 4.5 & 3 \\ \hline \hline 
\hline 
\end{tabular}
\caption{Effective Land\'e factors  $g=\frac{1}{\mu_B} \frac{dE}{dB}$ for B$\rightarrow$N1 transitions for the magnetic field oriented in $x$ (upper part) and $z$ direction (lower part of the table) as calculated from energy and transition spectra and the experimental results of Ref. \cite{lairdpei}. In the calculated results the derivative is taken at 0.05 T.
Due to a finite width of the transition peaks -- that are not strictly symmetric -- at the energy scale -- we estimate the precision in the evaluation of the $g$ factors to $\pm 0.1$. 
 }
\end{table}

\subsection{Detuning and the transition energy shifts}

The reduction of the $g$ factors discussed above has been obtained
as a result of the energy shifts of the transition lines that occur
for a relatively large amplitude of 5 kV/cm (potential drop of 0.5 mV along 100 nm), while in the experiment \cite{lairdpei} the amplitude of 0.5 kV/cm was applied. The factor which is crucial in the energy shifts discussed above
is the participation of the vacuum (0e,0h) state in the transitions,
and it varies not only with the amplitude but also on the bias $F_b$ between the dots.
The latter shifts the position of the (0e,0h) state
on the energy scale with respect to the (1e,1h) ground-state quadruple [Fig. \ref{widmoV}(b)].
The position of these two states is controlled in the EDSR experiment by detuning voltage applied between the dots \cite{lairdpei,extreme}.

\begin{figure}[htbp]
 \includegraphics[width=0.95\linewidth]{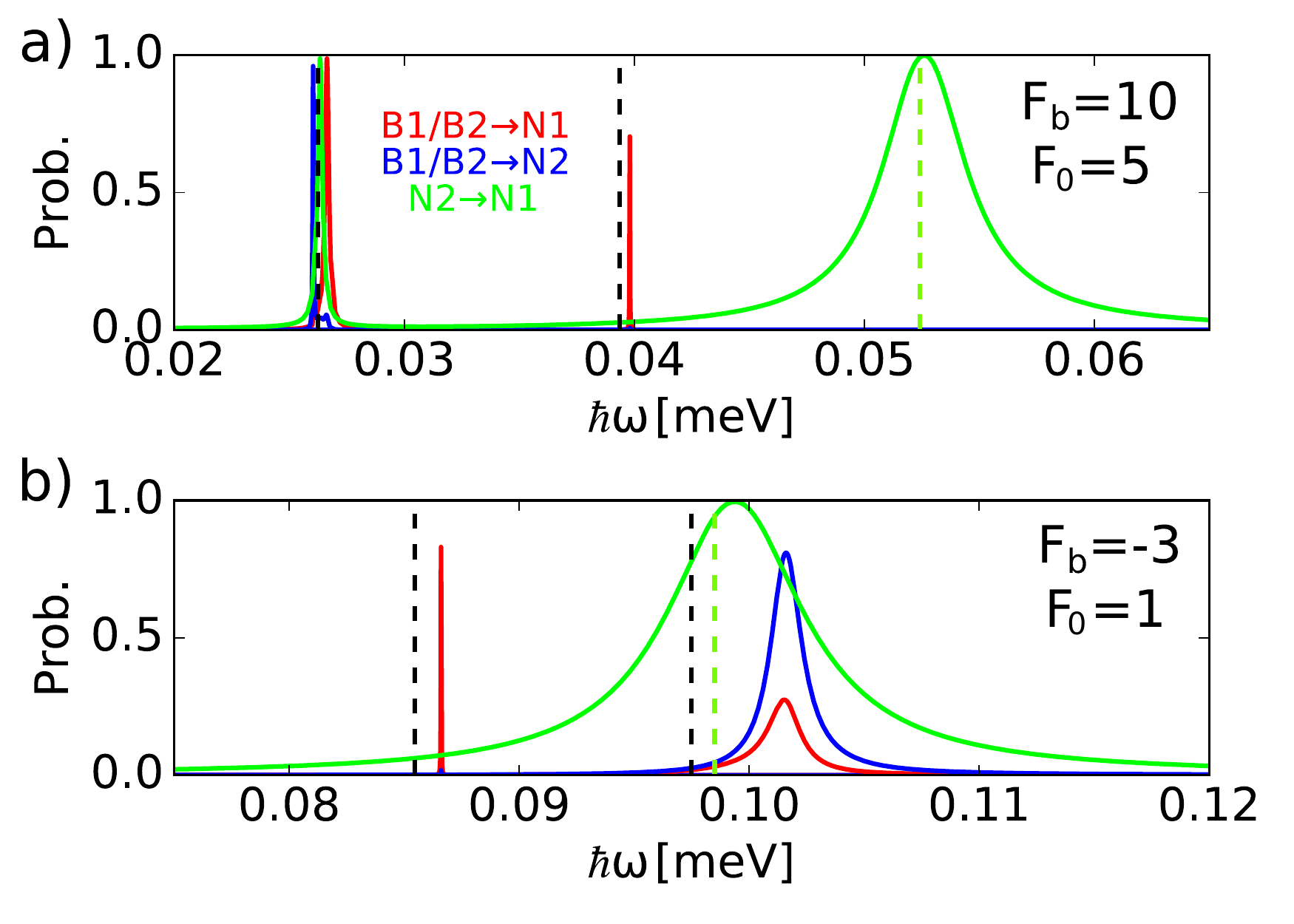} 
\caption{Maximal occupation probability obtained for a solution of the Schr\"odinger equation for 200 ns for $B_z=0.05$ T and (a) $F_0=5$ kV/cm, $F_b=10$ kV/cm, (b) $F_0=1$ kV/cm, $F_b=-3$ kV/cm. } \label{nununub}
\end{figure}

The role of the detuning for the energy shifts can be estimated from Figure \ref{nununub}(a),
where we plotted the result for $F_0=5$ kV/cm and $B_z=0.05$ (same as Fig. \ref{nunubasis}(d))
but for $F_b$=10 kV/cm. The (0e,0h) state is now about 40 meV above the (1e,1h) ground-state. The blue-shift of the transition peaks with respect to the energy splitting is reduced from large [Fig.8] to barely visible in 
Fig. \ref{nununub}(a). On the other hand -- for $F_0=1$ kV/cm -- for which the 
transition lines at $F_b=0$ coincide with the energy differences (see Fig. \ref{nunu0}(b) for $F_0=1$ kV/cm)  for $F_b=-3$ kV/cm [the (0e,0h) state $\simeq 17$  meV above the (1e,1h) ground-state], the blue shifts appear -- see Fig. \ref{nununub}(b). Concluding,
the transition energy shifts that stand behind the variation of the $g$ factors appear also for small amplitudes provided that the coupling with the (0e,0h) state is activated.

In the simulations the vacuum (0e,0h) state gets never very strongly occupied, and serves as a transition channel between the (1e,1h) energy levels of the lowest-energy quadruple. Nevertheless, the coupling between the vacuum (0e,0h) state and (1e,1h) states has a tunnelling character. 
The effects of the tunneling for a quantum wire -- quantum dot in the EDSR phenomenon was studied in Ref. \cite{Sherman}. The authors \cite{Sherman} found that the spin flipping  probability gets below 1  for stronger ac field and that the transition times are lower than expected for the Rabi oscillations. The first effect was encountered in Figs. 6(d) as according to the present study the effect  results from participation of a third
 state in the dynamics. We also reproduce the  other feature for larger ac fields. The B2$\rightarrow$ N1 transition time
for these two states included in the basis for the results of Fig. 8(a)  -- which produces the Rabi transition mechanism -- is 13 ns, while the spin-valley flip time for the convergent basis [Fig.8(e)] is 20 ns.

\subsection{Resonant transitions vs high harmonic generation}
The  results presented above contained a number of features
characteristic to nonlinear optics. Besides the shifts of the direct transition lines   also fractional resonances were observed, i.e. resonances at fractions of  the direct transition frequency (see Fig. \ref{nunubasis},\ref{nunu0},\ref{nunub}).
These transitions correspond to multiphoton absorption that is observed
in atomic optics at intense laser fields. For the gated nanodevices 
the conditions for observation of the phenomena characteristic to nonlinear optics \cite{extreme, li, pp}  appear at decent excitations of the order of 1 kV/cm (or the potential drop of 1 mV over 100 nm).  
In gaseous phase \cite{atto} and in solids \cite{aso}  strong laser fields ionize atoms and the ionized electrons 
are accelerated in the electric field of the laser. The oscillations of the dipole moment of the ionized electrons give rise to high harmonic generation \cite{lewen} that is used in generation of ultrashort pulses \cite{atto}. 

\begin{figure}[htbp]
 \includegraphics[width=0.95\linewidth]{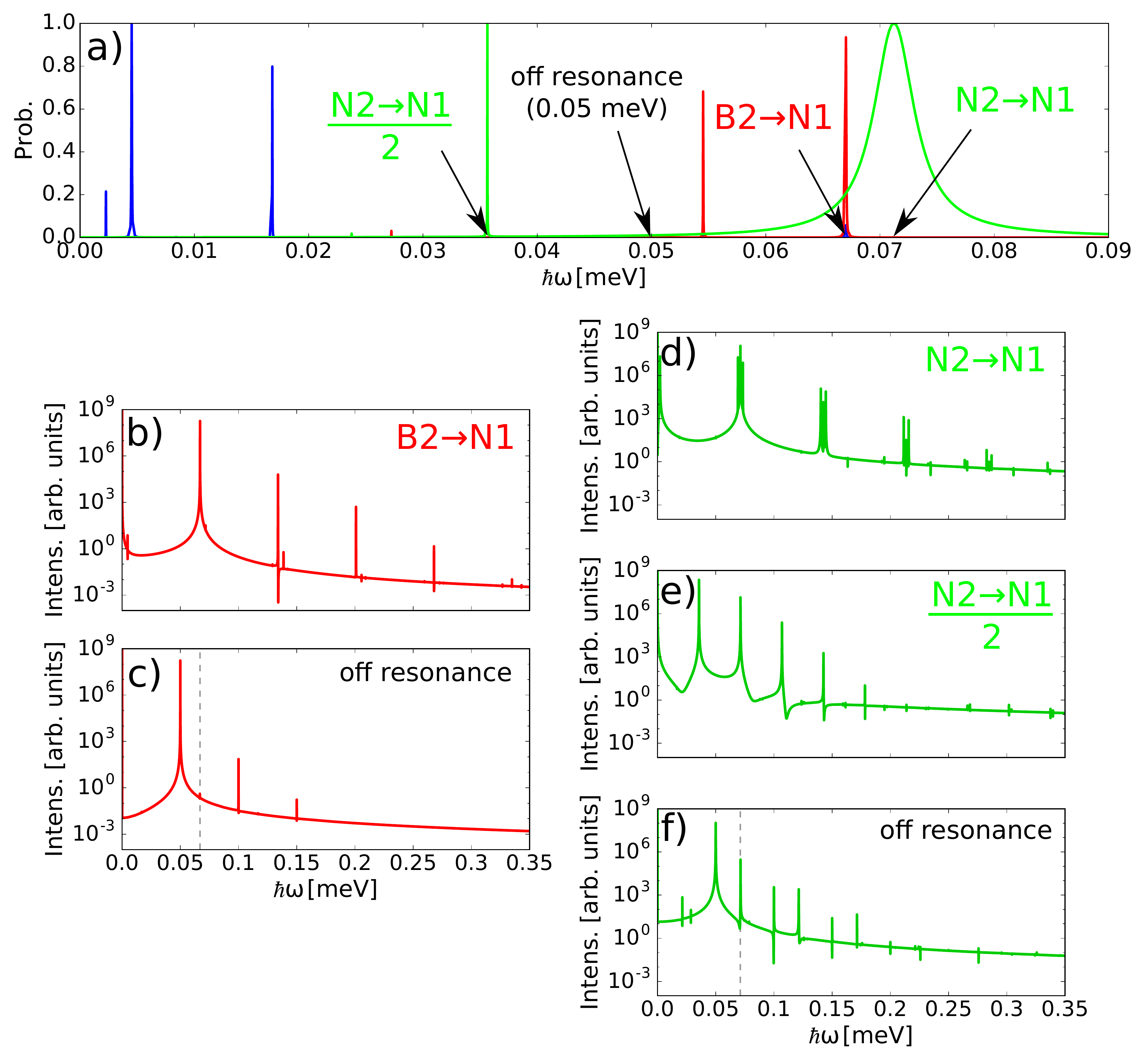}
\caption{Fourier transform of the electron dipole moment for $F_0=1$ kV/cm and $B_z=0.05$ T at no bias field. Panel (a) shows the driving frequencies $\hbar \omega_c$ considered in this plot. In (b) and (c) B2 is the
initial state. In (b) the driving frequency is $\hbar \omega_{ac}=0.067$ meV
-- resonant for the B2-N1 transition. In (c) $\hbar \omega_{ac}=0.05$ meV
we are off resonance. The dashed line in (b) shows the position of the resonant frequency of (b). In (d-f) the initial state is N2. In (d) and (e) resonant ac frequencies are set for the direct ($\hbar \omega_{ac}=0.0712$ meV) and two-photon ($\hbar \omega_{ac}=0.0356$ meV) N2-N1 transition. In (f) 
an off resonant driving frequency is applied ($\hbar \omega_{ac}=0.05$ meV).   The dashed line in (f) shows the position of the N2-N1 direct resonance frequency.
 } \label{furier}
\end{figure}

We looked for high harmonics of the electron dipole moments that are driven
by the ac field in our system \cite{lewen}. We considered $F_0=1$ kV/cm, $B_z=0.05$ T and no bias. Once the dynamics of the system is known we calculate the dipole moment $z_d (t) \equiv e\langle (z_1+z_2+z_3+z_4) \rangle$. Next, we Fourier analyze the dependence of the dipole moment on time. The results are presented in Fig. \ref{furier}. We set the initial state to B2 in Fig. \ref{furier}(b,c)
and N2 in Fig. \ref{furier}(d,e,f). The applied driving frequency of the electric field is taken resonant for the B2$\rightarrow$ N1 transition Fig. \ref{furier}(b) and resonant for the direct Fig. \ref{furier}(d)   and two-photon \ref{furier}(e) N2$\rightarrow$ N1 transitions  as well as off resonances [Fig. \ref{furier}(c,f)]. For resonant conditions we resolve up to 5th harmonic of the driving frequency [Fig. \ref{furier}(b,d,e)]. Off resonances the highest harmonic of the spectrum is the 3rd one [Fig. \ref{furier}(c,f)].

In off resonant conditions we notice  a peak
that correspond to the direct resonant transition B2-N1 [a tiny feature in Fig. \ref{furier}(c)] 
and N2-N1 [a pronounced feature in Fig. \ref{furier}(f)]. For the latter plot also other lines are observed. The N2-N1 transition line is wide at the energy scale and couples strongly to other transitions. 

Note, that in strong laser fields the high harmonic generation is a non-resonant phenomenon \cite{lewen,aso,atto}. In the present conditions the yield of the high harmonics is still strongly related to the resonant transition spectrum. 
In resonant conditions the 2nd harmonics is by one [Fig.\ref{furier}(e)], two [Fig.\ref{furier}(d)] or three orders [Fig.\ref{furier}(b)] of magnitude lower than the driving frequency. Off resonance the peak for the 2nd harmonic of the driving frequency is 4 [Fig.\ref{furier}(f)] or 6 [Fig.\ref{furier}(e)] orders of magnitude lower. 

Obviously, high harmonics generated in our systems have only moderate order 
in comparison to HHG generated in atoms  or molecules where the orders of 100 can be achieved \cite{Brabec-KraussRMP}. Still, the new mechanism discussed by us, when optimized could in principle lead to generation of ``truly high'' harmonics.

\section{Summary and conclusions}

We have analyzed spin-valley dynamics of the four last electrons in a n-p ambipolar quantum dot
using a time dependent configuration interaction method and the Floquet theory based on the single-electron states determined with the atomistic tight-binding approach. We studied the transitions lifting the Pauli blockade of the current within a quadruple of lowest-energy states of the (1e,1h) charge configuration.
We discussed  the results in the context of the accessible experimental data.

We demonstrated that the dynamics is significantly influenced by the coupling of the states of the (1e,1h) charge configuration with the nondegenerate vacuum (0e,0h) state. 
The vacuum state serves as a channel for transitions inside the (1e,1h) subspace
and its participation in the transitions is determined by both the amplitude of the ac electric field and the bias electric field.
The effect of the coupling are transitions energy shifts off the values
expected from the eigenenergy spectra. A strong modification of the $g$ factors characterizing the dependence of the transitions on external axial magnetic field. High harmonic generation in the electron dipole moment was found for resonant driving frequencies.

\section*{Acknowledgments}
This work was supported by the National Science Centre
according to decision DEC-2013/11/B/ST3/03837. 
E.N.O. benefits from the doctoral stipend ETIUDA of the National Science Centre
according to decision DEC-2015/16/T/ST3/00266 and the scholarship of Krakow
Smoluchowski Scientific Consortium from the funding for National
Leading Reserch Centre by Ministry of Science and Higher Education (Poland).
Calculations were performed in the PL-Grid Infrastructure.
M.L. and A.C. acknowledge Spanish MINECO grants (National Plan FOQUS No. FIS2013-46768-P and Severo Ochoa Excellence Grant No. SEV-2015-0522), the Catalan AGAUR grant SGR 874 2014-2016, and Fundaci\'o Privada Cellex Barcelona.

\end{document}


\author{E.N. Osika}
\affiliation{AGH University of Science and Technology, Faculty of Physics and
Applied Computer Science,\\
 al. Mickiewicza 30, 30-059 Krak\'ow, Poland}

\author{A. Chac\'on}
\affiliation{ICFO  -  Institut  de  Ciencies  Fotoniques,
The  Barcelona  Institute  of  Science  and  Technology,
08860  Castelldefels  (Barcelona),  Spain}

\author{M. Lewenstein}
\affiliation{ICFO  -  Institut  de  Ciencies  Fotoniques,
The  Barcelona  Institute  of  Science  and  Technology,
08860  Castelldefels  (Barcelona),  Spain}
\affiliation{ICREA, Pg. Llu\'is Companys 23, 08010 Barcelona, Spain}

\author{B. Szafran}
\affiliation{AGH University of Science and Technology, Faculty of Physics and
Applied Computer Science,\\
 al. Mickiewicza 30, 30-059 Krak\'ow, Poland}

\title{Spin-valley dynamics of electrically driven ambipolar carbon-nanotube quantum dots - Supplementary material}
\begin{abstract}

\end{abstract}
\maketitle

\global\long\def\thefigure{S\arabic{figure}}
This supplementary material contains additional analysis of the role of the vacuum state (0e,0h) and the  emergence of B2 $\rightarrow$ N2 transition 
for ac driving frequency to B2 $\rightarrow$ N1 resonance for $B_z=0.05$ T and $F_0=5$ kV/cm -- as in Fig. 6(d) of the paper.
We limited the basis to the 5 states: {N1,B1,B2,N2,(0e,0h)} that govern the dynamics of the system [see the convergence test of Fig. 9 in the paper].
In Fig. \ref{czas_zoom} we repeat a very beginning of the simulation presented in Fig. 6(d) only using the limited basis. We plot  
the vacuum state (0e,0h) occupation -- previously omitted. The occupation of the vacuum state 
appears in peaks which are followed by a step of the occupation of the N2 state.

\begin{figure}[htbp]
 \includegraphics[width=0.9\linewidth]{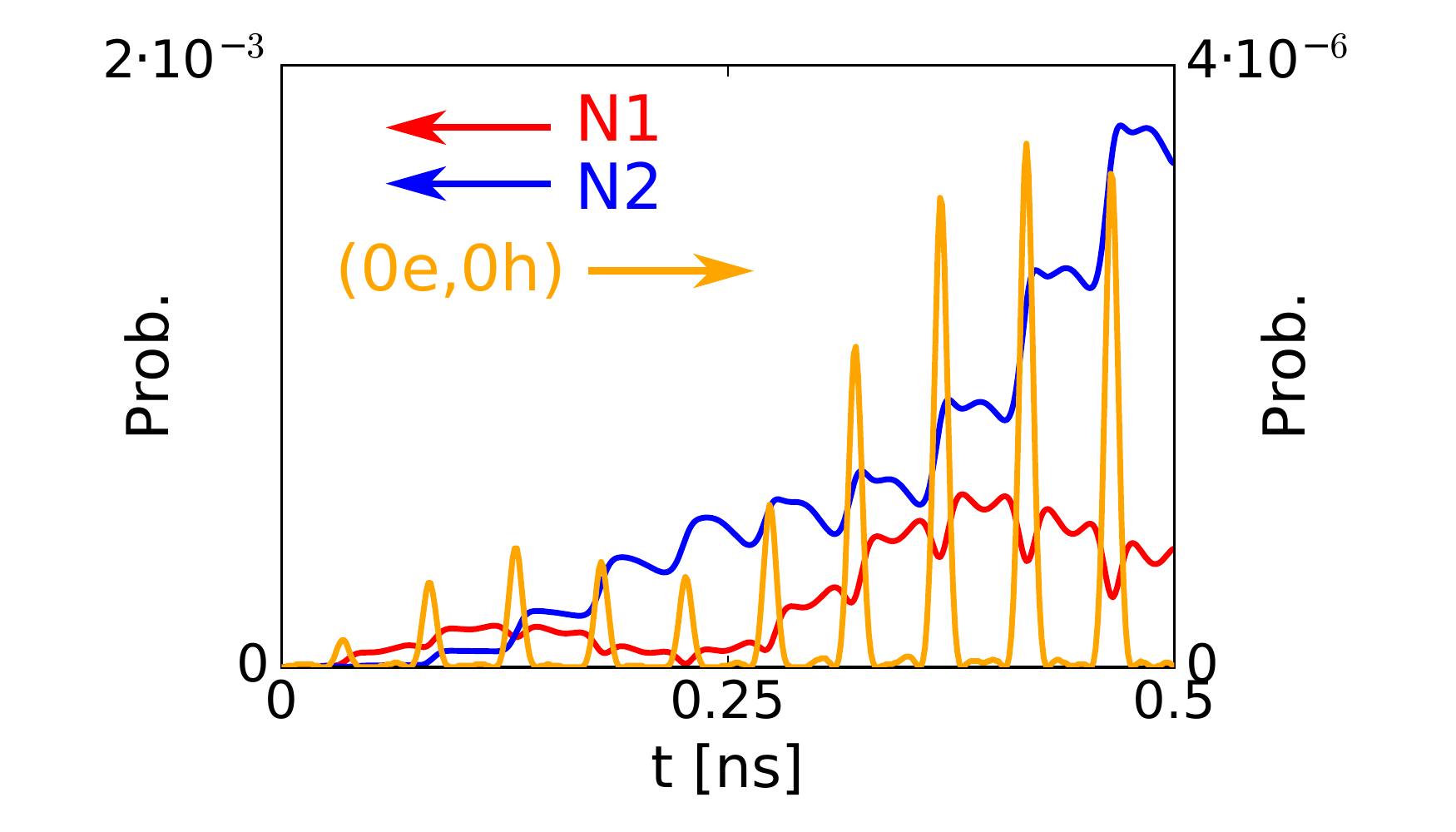}
 \caption{ N1, N2 and (0e,0h) occupation probability at the beginning of the B2$\rightarrow$N1 resonant transition at $B_z=0.05$ T and $F_0=5$ kV/cm and the limited basis.
} \label{czas_zoom}
\end{figure}

The role of the vacuum state  (0e,0h)  in the dynamics of the system can be followed by an analysis of the solution
of the Schr\"{o}dinger equation in the configuration interaction method
\begin{equation}
i\hbar \dot{c}_k(t)=\sum_n c_n(t) eF_0 \sin(\omega t) \langle\Psi_k|z|\Psi_n\rangle e^{-\frac{i(E_n-E_k)t}{\hbar}} \label{cndot}.
\end{equation}
We integrate both sides over time with limits $0$ and $\tau$ and obtain
\begin{eqnarray}
 &c_k(\tau)&=c_k(0) + \nonumber \\
 &\frac{1}{i\hbar}\int\limits_0^\tau & \sum_n c_n(t) eF_0 \sin(\omega t) \langle\Psi_k|z|\Psi_n\rangle e^{-\frac{i(E_n-E_k)t}{\hbar}} dt \equiv \nonumber \\
 &c_k(0)&  + \sum_n J_{kn}(\tau) \label{cndot2},
\end{eqnarray}
 The integrated matrix elements $J_{kn}(\tau)$ contain information on the transitions between the basis states. 

In Fig. S2 we label the five states of the limited basis [N1,B1,B2,N2,(0e,0h)] by their energy order in the spectrum as 1-st,2-nd,3-rd,4-th and 17-th state in the basis, respectively.
In the initial condition the system occupies state B2 (i.e. the 3rd one ). The third column of Fig. S2 indicates 
that the driving -- resonant with B2$\rightarrow$N1 transition (or $3\rightarrow 1$) induces most effectively
the transition to the vacuum state (0e,0h) (17-th). The transition flux $J_{17,4}$ is about 30 times more effective than $J_{1,4}$. 
The last column of Fig. S2 shows that transitions from the vacuum state go to both N1 (1st) and N2 (3rd) states. 
The direct transitions between the N1 and N2 states are by 1.5 or 2 more effective than N$\leftrightarrow$(0e,0h) transitions
- cf. $J_{4,1}$ and $J_{4,17}$ and $J_{17,4}$ and $J_{17,1}$, etc.


\begin{figure*}[htbp]
 \includegraphics[width=0.9\linewidth]{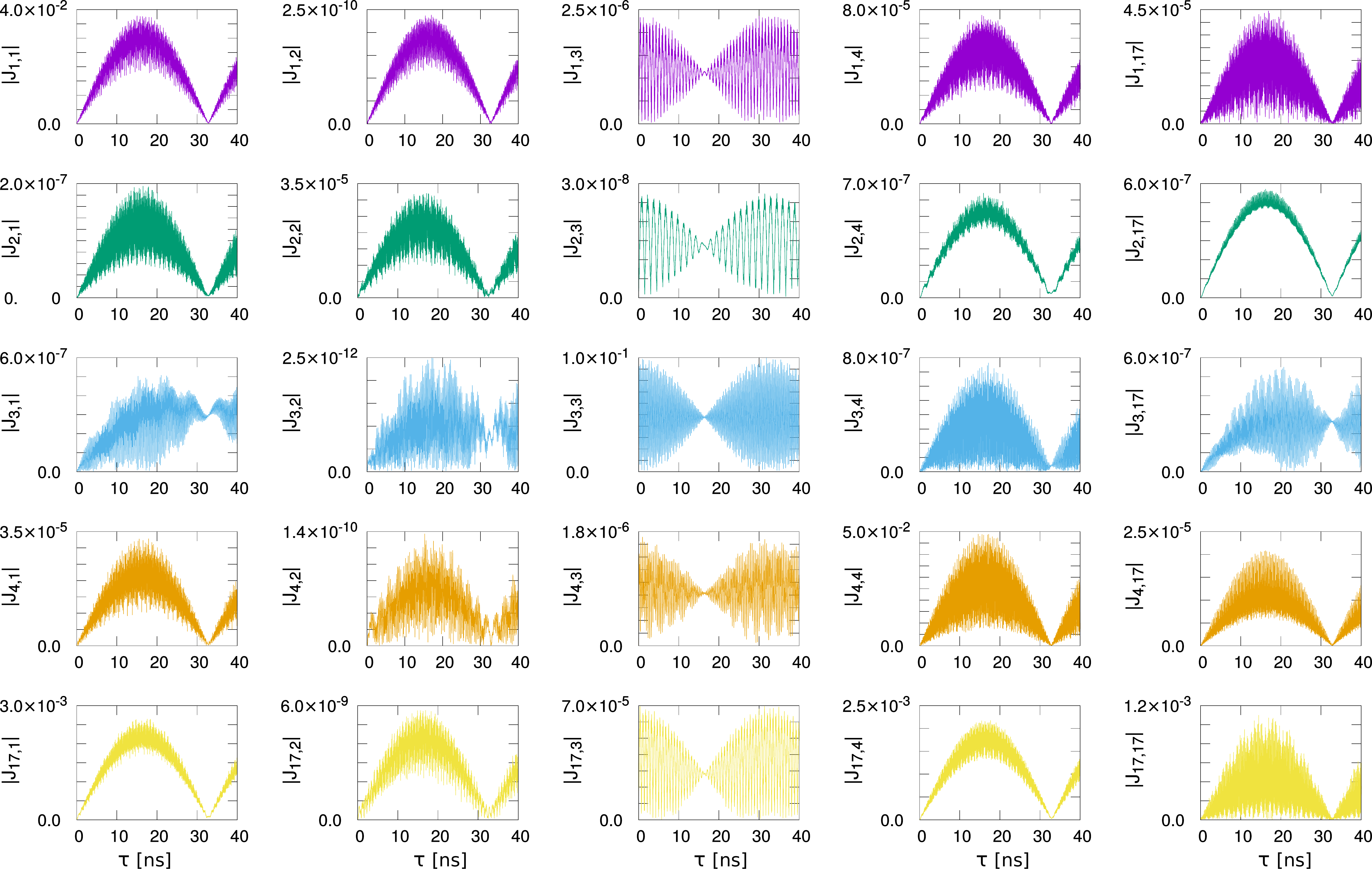}
 \caption{ $|J_{kn}|$ elements as a function of $\tau$ for the driving frequency set in resonance to  B2$\rightarrow$N1 transition at $B_z=0.05$ T and $F_0=5$ kV/cm.
} \label{cint_abs}
\end{figure*}